\documentclass[aps,prapplied,twocolumn,superscriptaddress]{revtex4-2}
\usepackage{graphicx}
\usepackage{amssymb,amsfonts,amsmath}
\usepackage[colorlinks=true,citecolor=Cerulean,linkcolor=RubineRed,urlcolor=Cerulean]{hyperref}
\hypersetup{breaklinks=true}
\usepackage{graphicx}
\usepackage{color}
\usepackage[usenames,dvipsnames]{xcolor}
\usepackage{epstopdf}
\usepackage[normalem]{ulem}
\usepackage{bm}
\usepackage{bbold}
\usepackage{float}
\usepackage{dsfont}

\renewcommand{\d}{\mathrm{d}}

\newcommand{\bv}[1]{{\boldsymbol #1}}
\renewcommand{\i}{{\rm i}}
\newcommand{\ket}[1]{|#1\rangle}
\newcommand{\bra}[1]{\langle #1|}
\newcommand{\ketbra}[2]{|#1\rangle \langle #2 |}
\renewcommand{\vec}[1]{{\bf #1}}

\begin{document}

\title{Deterministic Creation of Large Photonic Multipartite Entangled States with Group-IV Color Centers in Diamond}

\author{Gregor~Pieplow}
\affiliation{Department of Physics, Humboldt-Universität zu Berlin, 12489 Berlin, Germany}%

\author{Yannick Strocka}
\affiliation{Department of Physics, Humboldt-Universität zu Berlin, 12489 Berlin, Germany}%

\author{Mariano~Isaza-Monsalve}
\affiliation{Department of Physics, Humboldt-Universität zu Berlin, 12489 Berlin, Germany}%
\affiliation{Institute for Experimental Physics, Universität Innsbruck, 6020 Innsbruck, Austria}

\author{Joseph~H.~D.~Munns}
\affiliation{Department of Physics, Humboldt-Universität zu Berlin, 12489 Berlin, Germany}%

\author{Tim~Schr{\"o}der}
\affiliation{Department of Physics, Humboldt-Universität zu Berlin, 12489 Berlin, Germany}%
\begin{abstract}
Measurement-based quantum computation relies on single qubit measurements of large multipartite entangled states, so-called lattice-graph or cluster states. Graph states are also an important resource for quantum communication, where tree cluster states are a key resource for one-way quantum repeaters. A photonic realization of this kind of state would inherit many of the benefits of photonic platforms, such as very little dephasing due to weak environmental interactions and the well-developed infrastructure to route and measure photonic qubits. In this work, a linear cluster state and GHZ state generation scheme is developed for group-IV color centers. In particular, this article focuses on an in-depth investigation of the required control operations, including the coherent spin and excitation gates. We choose an off-resonant Raman scheme for the spin gates, which can be much faster than microwave control. We do not rely on a reduced level scheme and use efficient approximations to design high-fidelity Raman gates. We benchmark the spin-control and excitation scheme using the tin vacancy color center coupled to a cavity, assuming a realistic experimental setting. Additionally, the article investigates the fidelities of the Raman and excitation gates in the presence of radiative and non-radiative decay mechanisms. Finally, a quality measure is devised, which emphasizes the importance of fast and high-fidelity spin gates in the creation of large entangled photonic states. 
\end{abstract}
\maketitle

\section{Introduction}
    
    Large, multipartite entangled photonic states, so-called photonic lattice-graph states or cluster states (CS), are an integral resource in two key quantum information applications: measurement-based quantum computing \cite{raussendorf_one-way_2001, walther_experimental_2005, obrien_optical_2007, Briegel2009}, and one-way quantum repeaters \cite{azuma_all-photonic_2015, buterakos_deterministic_2017, borregaard_one-way_2020, wo_resource-efficient_2023}.   
    There are several approaches to generating these resource states: by using optical parametric oscillators \cite{menicucci_one-way_2008, yokoyama_ultra-large-scale_2013, larsen_deterministic_2019} (continuous variable CS), by combining entangled photon pair states \cite{lu_experimental_2007} generated with spontaneous parametric down conversion \cite{walther_experimental_2005, yao_experimental_2012}, by using ultra cold atoms \cite{thomas_efficient_2022}, and by using on-demand solid-state emitters such as quantum dots \cite{lindner_proposal_2009, schwartz_deterministic_2016, lee_quantum_2019, cogan_deterministic_2023} or nitrogen vacancy centers (NVs) \cite{rao_generation_2015, vasconcelos_scalable_2020}.
    
    This work focuses on the theoretic investigation of a new platform for the generation of CS: group-IV color centers in diamond (G4Vs) \cite{goss_twelve-line_1996, thiering_ab_2018} including the silicon vacancy (SiV), the germanium vacancy (GeV), the tin vacancy (SnV), and the lead vacancy (PbV). More specifically, this work provides a detailed analysis of spin-photon entanglement and the coherent control needed for generating single spin qubit gates, which are both essential for the deterministic generation of CS through emission-based schemes  \cite{lindner_proposal_2009, schwartz_deterministic_2016, lee_quantum_2019, tiurev_fidelity_2021, thomas_efficient_2022, tiurev_high-fidelity_2022, cogan_deterministic_2023}. 
    \begin{figure*}[ht!]
        \centering
        \includegraphics[width = \textwidth]{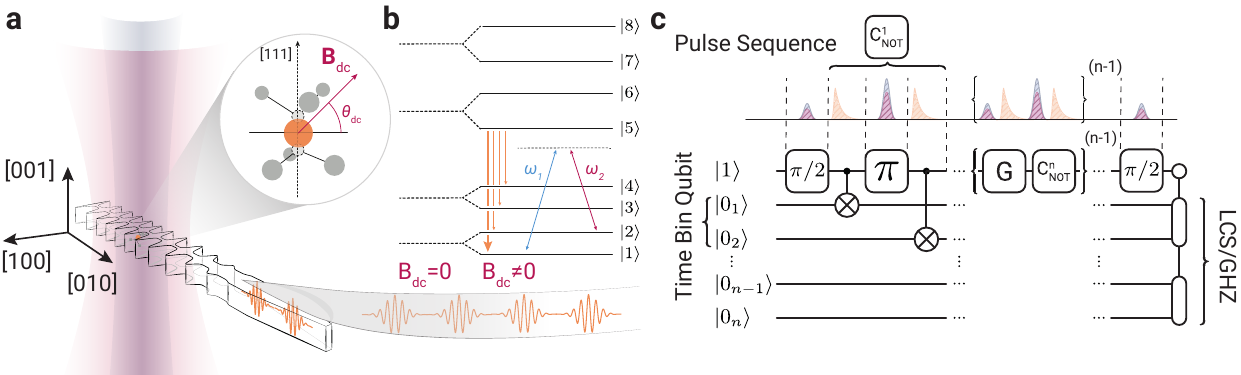}
        \caption{a) Sawfish cavity \cite{bopp_sawfish_2022} creating Purcell enhancement of the desired transition indicated by the thick orange arrow in the level scheme (b). The control laser fields, which implement the spin gates, are indicated by the blue and magenta cones, which interact with a G4V at the center of the cavity. Time-bin entangled photons (orange) are emitted into a waveguide and subsequently into a fiber. b) Level scheme of the G4V in the presence of a magnetic field, lifting the spin degeneracy. The laser's carrier frequencies are indicated by $\omega_1$ and $\omega_2$. The thick orange arrow represents the Purcell enhanced transition. In principle, decay into the other spin levels is allowed (thin orange arrows), which is detrimental to the CS generation. c) Emission-based linear cluster state generation scheme (circuit representation and corresponding pulse sequence). The system is initialized in $|1\rangle$ and then prepared in an equal superposition $(|1\rangle + |2\rangle)/\sqrt{2}$ with an all-optical $\pi/2$ rotation (blue and red pulses) laser pulse. The $|1\rangle \rightarrow|5\rangle$ transition is then driven resonantly (light orange pulse), producing a photon with 50\% probability in the early time-bin $|e_1\rangle = |1_1\rangle|0_2\rangle$. A $\pi$-rotation (red and blue pulses) is then followed by another resonant pulse producing a spin and photonic time-bin qubit entangled state (CNOT-Gate) $(|2\rangle|e_1\rangle- |1\rangle|l_1\rangle)/\sqrt{2}$, where $|l_1\rangle = |0_1\rangle|1_2\rangle$. Repeating the procedure with $G = \pi/2$ produces a linear cluster state (LCS) of time-bin entangled photons, which can be detached by a Z-measurement of the spin qubit. In the case of $G = \mathbb{1}$, a Greenberger-Horne-Zeilinger state (GHZ) is produced.}
        \label{fig:fig0}
    \end{figure*}
    The biggest challenges in deterministic CS generation with a solid state emitter are maintaining coherence, interfacing with single coherent photons, and minimizing photon losses \cite{tiurev_fidelity_2021}. Due to their properties, G4Vs in diamond are ideal candidates to address these challenges: G4Vs have been shown to exhibit competitive $T_2$ spin coherence times. Specifically, for the SiV a $T_2 \approx 13$ ms at 100 mK \cite{sukachev_silicon-vacancy_2017} has been achieved. Due to the larger orbital ground state splitting, the spin coherence time of the heavier G4V such as the SnV and PbV is typically longer \cite{iwasaki_tin-vacancy_2017} compared to the SiV and GeV at temperatures below $4$ K \cite{harris_coherence_2023}, making it feasible to achieve ms coherence times with liquid helium cryostats. In particular, for the SnV $T_2 \approx 0.3~$ms at $1.7~$K \cite{debroux_quantum_2021} has been demonstrated. For this reason, when characterizing the performance of CS generation, numerical estimates are obtained for the SnV as a case study.
    
    G4V have been shown to emit transform-limited photons \cite{arend_photoluminescence_2016, trusheim_transform-limited_2020, chen_quantum_2022}, which is key for the creation of indistinguishable single photons. 
    Compared to the NV's Debye-Waller factor of ${\rm DWF= 3\%}$ \cite{faraon_resonant_2011, alkauskas_first-principles_2014}, G4Vs also posses a notably larger DWF between 60\% and 80\% \cite{siyushev_optical_2017,lindner_strongly_2018,gorlitz_spectroscopic_2020}.    
    Additionally, due to their inversion symmetry \cite{bradac_quantum_2019}, G4Vs are much less susceptible to spectral diffusion caused by charge noise \cite{rogers_multiple_2014, sipahigil_indistinguishable_2014, gorlitz_spectroscopic_2020} in contrast to the NV \cite{faraon_coupling_2012, orphal-kobin_optically_2023}.
    Their reduced sensitivity to charge noise, which is exacerbated close to surfaces in, for example, integrated structures such as photonic crystal nanocavities, makes G4Vs a particularly appealing platform for CS generation \cite{schroder_scalable_2017, rugar_quantum_2021}. In a recent proposal \cite{bopp_sawfish_2022}, the integration of G4V into highly efficient nanophotonic crystal cavities, the so-called sawfish cavity \cite{pregnolato_fabrication_2023} (see Fig.~\ref{fig:fig0}a) was simulated. The sawfish design allows for almost unit efficiency coupling of the emission of a single emitter into an optical fiber. Cavities are required for the generation of CS in either emission- or scattering-based schemes \cite{shi_deterministic_2021}. They facilitate the enhanced coupling of the emitter to the cavity mode \cite{bhaskar_experimental_2020}, which can be used for spin-dependent reflection and therefore spin-photon entanglement \cite{shi_deterministic_2021} or for reducing branching errors by shortening the lifetime of a desired transition in emission-based schemes \cite{tiurev_fidelity_2021}.
    
    In this work, we benchmark the cavity enhanced cluster state creation process using two multipartite entangled states: a linear cluster state (LCS) and the Greenberger-Horne-Zeilinger (GHZ) state. The protocol for their generation is shown in Fig.~\ref{fig:fig0}c. More precisely, we analyze four important steps in their creation, which are $\pi/2$ and $\pi$ rotation gates operating on the spin qubit, the resonant spin-dependent excitation producing a single photon, and the subsequent radiative decay producing time-bin entangled photons. 
    
    The single qubit rotation gates are realized by an all-optical off-resonant Raman scheme \cite{chen_theory_2004, caillet_precision_2007,vasilev_optimum_2009, campbell_ultrafast_2010, vitanov_stimulated_2017}. Such schemes have been proposed \cite{chen_theory_2004} and implemented \cite{press_complete_2008} for quantum dots, atomic systems \cite{yavuz_fast_2006, campbell_ultrafast_2010}, and G4Vs \cite{becker_ultrafast_2016, Debroux2021}. 
    The Raman scheme uses two laser pulses (see Fig. \ref{fig:fig0}b) that couple the states $|1\rangle$ and $|2\rangle$ in the lowest lying orbital branch through the excited state $|5\rangle$. 
    
    The scheme doesn't rely on complex pulse shaping \cite{lacour_implementation_2006, takou_optical_2021} and is compatible with ultrafast gates of tens of picoseconds \cite{becker_ultrafast_2016}. Such short gate times are out of the reach for most feasible microwave control gates, which are tens of nanoseconds long \cite{pingault_coherent_2017}. Fast gates are especially important when multiple photons are sequentially entangled, due to both the dephasing of the emitter and the photon loss in the transmitting fiber. An all-optical scheme is also better suited to address individual qubits in nanostructures due to the shorter wavelengths of visible light. Furthermore it can bridge ground-state splittings that are much larger than those addressable by microwaves. 
    
    Here we analytically and numerically include the far-off-resonance excited states, which can introduce small errors when unaccounted for. We also account for the directionality of the magnetic field, which lifts the spin degeneracy in the design of spin gates, and we include its orientation in the engineering of ultra high-fidelity spin gates. We perform the optimization in the presence of photonic and phononic decay processes, which we identified as the most obstructive intrinsic processes for the preparation of high fidelity gates. Finally, we evaluate the control sequences in the context of the creation of a linear cluster state. For this purpose, we introduce an original time-bin multipartite entanglement quality measure, which underlines the importance of quick and high-fidelity control techniques when creating large time-bin entangled photonic qubit states. 
    
    This article is structured as follows: First, in section \ref{section:cs_gen} the deterministic LCS generation protocol with G4Vs based on the time-bin entanglement scheme is introduced. Its main features are described and processes that affect the spin gates and therefore the state fidelities are discussed. In section \ref{sec:controlop} the general Raman control framework and a gate optimization procedure is presented, which is then applied to both spin and excitation gates. In section \ref{section:cluster} we introduce a linear cluster state quality measure and the previously obtained gate fidelities are evaluated in the context of linear CS creation. In section \ref{section:discussion}, the main results of our work are discussed.    
    
\section{Linear Cluster and GHZ state generation \label{section:cs_gen}}
     
    Here we present the idealized scheme for generating a LCS and a GHZ state with G4Vs coupled to a cavity. The protocol is rooted in the emission-based protocol introduced in \cite{lee_quantum_2019}. Fig. \ref{fig:fig0}c shows the protocol in a circuit representation. The relevant spin states of the spin 1/2 system that is constituted by a negatively charged G4V \cite{hepp_electronic_2014, thiering_ab_2018} are shown in Fig. \ref{fig:fig0}b. A magnetic field $\vec{B} = B [\cos(\theta_{\rm dc}), 0, \sin(\theta_{\rm dc})]$, which is not aligned with the symmetry axis of the color center, is required to lift the spin degeneracy and allows for the coupling of anti parallel spin states through the Raman control fields. 
    
    The system is first initialized in $\ket{1}$ ($\ket{1}$, $\ket{2}$ constitute the spin qubit). 
    Two Raman pulses (shown as the red and blue pulse in Fig. \ref{fig:fig0}) realize a $\pi/2$ rotation around the $y$-axis on the Bloch sphere, so that $U_{\pi/2} = \exp(-\i \sigma_y \pi/4)$, where $\sigma_y$ is the y-Pauli matrix, which results in an equal superposition $(\ket{1} + \ket{2})/\sqrt{2}$ . 
    The $\ket{1} \leftrightarrow \ket{5}$ transition is then driven by two resonant excitation pulses (light orange pulses), which are separated by a spin $\pi$-rotation $U_{\pi} = \exp(-\i \sigma_y \pi/2)$ (red and blue Raman pulse). 
    The two resonant excitation pulses and the $\pi$-rotation create a single photon that is emitted with a $50\%$ probability into either an early $\ket{e_1}$ or late $\ket{l_1}$ time-bin, which together constitute the photonic time-bin qubit. This pulse sequence entangles the electron spin with the photonic time-bin qubit: $(\ket{2}\ket{e_1} -\ket{1}\ket{l_1})/\sqrt{2}$ \cite{lee_quantum_2019, tchebotareva_entanglement_2019}. The CNOT gate in Fig. \ref{fig:fig0} is composed of the pulse sequence consisting of the two excitations and the $\pi$ rotation. 
    Repeating these steps ($G = \pi/2$ and CNOT gate) $n-1$ times creates a LCS of length $n-1$, which can be detached from the spin-qubit after the last $\pi / 2$ rotation by a Z-measurement (for more details see \cite{lee_quantum_2019}). If $G = \mathbb{1}$ instead, a GHZ state is created. Ideally the excitation produces $\ket{1} + \ket{2} \rightarrow e^{\i \phi}\ket{5} + \ket{2}$, where $e^{\i \phi}$ contributes to a global phase, due to the double excitation in each entangling step.  

\section{Control operations}\label{sec:controlop}

    Here we present the control steps that are required in the generation of a LCS, a GHZ state, and their anticipated fidelity. They consist of the coherent Raman control scheme of the spin qubit, the resonant excitation, and the spontaneous radiative decay, which sequentially produces the time-bin entangled photons.  

    \subsection{Coherent Raman Control \label{section:control}}
        \begin{figure}
            \centering
            \includegraphics[width=\columnwidth]{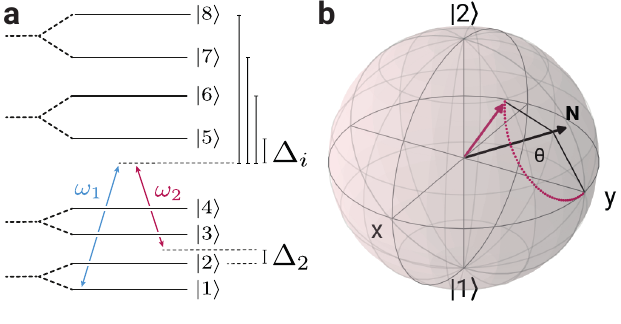} 
            \caption{a) Level scheme of a G4V in the presence of a magnetic field $\vec{B} = B [\cos(\theta_{\rm dc}), 0, \sin(\theta_{\rm dc})]$ (no strain). The carrier frequencies of the two laser fields are indicated by $\omega_1$ and $\omega_2$. The magnetic field has to be of axis for optical Raman control, because otherwise Raman rotations of $|1\rangle$ and $|2\rangle$ would fail. The detunings are $\Delta_2 = \epsilon_2-\epsilon_1 - (\omega_1-\omega_2)$ and $\Delta_{i} = \epsilon_i - \epsilon_1 - \omega_1$, $i>4$ (see \ref{appsec:rwa}). b) Raman rotation on the Bloch sphere \eqref{eq:bloch_rot}, where the components of the direction of the rotational axis $\vec{N}$ are fixed by \eqref{eq:rot-axis} and the angle $\theta$ by \eqref{eq:rot-theta1}}
            \label{fig:fig1}
        \end{figure}
        In this section, the off-resonant Raman control scheme for performing high fidelity rotations within the G4V ground state manifold is presented. Raman control has already been proposed and implemented for G4V vacancies \cite{becker_ultrafast_2016, Debroux2021} but, interestingly, single pulse $\pi$-rotations have not yet been experimentally demonstrated. We are also not aware of any in-depth study of the off-resonant Raman control scheme of G4V presented in this work: here we emphasize the need to take into account the entire level structure and the details of the magnetic field when designing fast, high fidelity control gates for G4Vs.
        
        We first derive approximate control parameters, such as the electric field strengths of the laser pulses, then we perform an optimization procedure. 
        The G4V is driven by a superposition of two classical laser pulses with central frequencies $\omega_1$ and $\omega_2$
        \begin{align}
            \begin{aligned}
            \vec{E}(t)=  &E_1(t)(\vec{e}_1e^{\i \omega_1 t} + \vec{e}_1^*e^{-\i \omega_1 t})/2 
            \\
            &+ E_2(t)(\vec{e}_2e^{\i \omega_2 t} + \vec{e}_2^*e^{-\i \omega_2 t})/2,
            \end{aligned}
            \label{eq:e-field-time-dep}
        \end{align}
        where $E_{1,2}(t)$ is a real valued time dependent envelope that changes on time scales that are much slower compared to $2 \pi/\omega_{1,2}$, where $\omega_{1,2}$ are the lasers pulses' carrier frequencies. 
        The complex polarization vectors $\vec{e}_{1,2}$ have unit length and encode the relative phase of the two laser pulses.
        The interaction of the classical driving field with the color center is given by the electric dipole interaction
        \begin{equation}
            H(t) =  H_0-\bv{\mu} \cdot \vec{E}(t),
            \label{eq:spin_h}
        \end{equation}
        where $H_0 = \sum_{i=1}^8 \epsilon_i |i\rangle\langle i|$ is the Hamiltonian expanded in the eigenstates $|i\rangle$ of 
        \begin{equation}
            H_0 = H_{\rm A}+H_{\rm SO} + H_{\rm JT} + H_{\rm Z}~.
        \end{equation}
        $H_{\rm A}$ is the color centers' bare Hamiltonian with two four-fold degenerate eigenstates with energies $\epsilon_{g,u}$, $H_{\rm SO}$ is the spin-orbit interaction, $H_{\rm JT}$ the Jahn-Teller and $H_{\rm Z}$ the Zeeman contribution (see Appendix~\ref{appsec:hamiltonian}). The corresponding level scheme in relation to $\omega_{1,2}$ is shown in Fig.~\ref{fig:fig1}. In this work we do not consider strain \cite{pieplow_efficient_2023}. The energies $\epsilon_i$ are in units of angular frequency. 
        The dipole transition vector is 
        \begin{equation}
            \bv{\mu} = (\mu^x, \mu^y, \mu^z)~,
        \end{equation}
        and its components expanded in the orbital basis are 
        \begin{equation}
            \mu^\alpha = \sum_{i,j=1}^8 \mu_{ij}^\alpha |i\rangle\langle j|
            \label{eq:spin-mu}
        \end{equation}
        There are no dipole allowed transitions within either the ground or excited state manifolds \cite{hepp_electronic_2014}.
        
        Now we calculate approximate parameters for generating the $\pi/2$ and $\pi$ rotations around the $y$-axis that are required for cluster state generation. 
    
        In a rotating frame, the Hamiltonian \eqref{eq:spin_h} can be written as (Appendix \ref{appsec:rwa}) 
        \begin{equation}
            H'(t) = H_{\rm RWA}(t) + H_{\rm F}(t) + H_{\rm fast}(t)~,
            \label{eq:ham-rot-frame}
        \end{equation}
        where $ H_{\rm RWA}(t)$ inherits its time dependence from the slowly varying amplitudes $E_k(t)$ [Eq.\eqref{eq:e-field-time-dep}]. 
        The Hamiltonian $H_{\rm F}(t)$ is quasi periodic in time and features oscillations at the frequency  $\omega_1 - \omega_2$ and 
        the Hamiltonian $H_{\rm fast}(t)$ contains terms oscillating at frequencies $2 \omega_{1,2}$ as well as $\omega_1 + \omega_2$. 
        
        Neglecting the terms $H_{\rm F}(t)$ and $H_{\rm fast}(t)$ corresponds to the rotating wave approximation (RWA).
        When only neglecting $H_{\rm fast}(t)$, we refer to the Floquet approximation.
        
        After performing the rotating wave approximation and adiabatically eliminating the excited states (Appendix~\ref{appsec:rwa}), the effective time evolution of the spin qubit $|1\rangle$ and $|2\rangle$ is generated by the Hamiltonian
        \begin{align}
            H_{\rm eff}(t) = \vec{n}(t) \cdot \bv{\sigma} + \Delta_{\rm eff}(t)/2 ~,
            \label{eq:eff-bloch-rot}
        \end{align}
        where $\bv{\sigma} = (\sigma_x,\sigma_y,\sigma_z)$, $\sigma_\alpha$ are the Pauli matrices,
        \begin{equation}
            \vec{n}(t) = [\Omega_{\rm eff}'(t), \Omega_{\rm eff}''(t), -\Delta_{\rm eff}(t)/2]~,
            \label{eq:rot-axis}
        \end{equation}
        and $\Omega_{\rm eff} = \Omega_{\rm eff}' + \i\Omega_{\rm eff}''$ is given by 
        \begin{align}
            &\Omega_{\rm eff}(t) = \sum_{i=5}^8 \frac{\Omega_{1i; 1}(t)\Omega_{2i; 2}(t)^*}{\Delta_{i}}
           \\
           &\Delta_{\rm eff}(t) = \Delta_2 + \sum_{i=5}^8\frac{1}{\Delta_i}\left(|\Omega_{2i,2}(t)|^2 -|\Omega_{1i,1}(t)|^2 \right)
        \end{align}
        where
        \begin{equation}
            \Omega_{ij,k}(t) = -\frac{1}{2}\bra{i}\bv{\mu}\cdot \vec{e}_k\ket{j} E_k(t)~,
            \label{eq:coupling}
        \end{equation}
        and $\Delta_{i} = \epsilon_{i} - \epsilon_1  - \omega_1$ for $i>4$ and the two-photon detuning is given by $\Delta_2 = \omega_2 - \omega_1 - \epsilon_2$.
        
        If the laser pulses have the same shape, a finite duration and interact simultaneously with the color center such that 
        \begin{equation}
            E_{1,2}(t) = \mathcal{E}_{1,2} \xi(t)~,
            \label{eq:envelope}
        \end{equation} and $\Delta_2 = 0$ then  
        \begin{align}
            \begin{aligned}
                \vec{n}(t)\cdot \bv{\sigma} \rightarrow \vec{n}\cdot\bv{\sigma} \xi(t)^2~. 
            \end{aligned}
        \end{align}
        The interaction of the color center with the two laser pulses is then given by the unitary operator 
        \begin{equation}
            R_\theta =  e^{\i \Phi}e^{-\i\theta \vec{N}\cdot\bv{\sigma}/2}~,
        \label{eq:bloch_rot}
        \end{equation}
        where $\Phi$ is a global phase that is neglected in the following, $\vec{N}=\vec{n}/|\vec{n}|$ and
        \begin{equation}
            \theta = 2 \sqrt{|\Omega_{\rm eff}|^2 + \Delta_{\rm eff}^2} \int_{-\infty}^{\infty}\!\!\!\d t\, \xi^2(t) / \hbar 
            \label{eq:rot-theta1}.
        \end{equation}
        The unitary operator $R_\theta$ can be interpreted as a rotation around $\vec{N}$ by an angle $\theta$ on the Bloch sphere (see Fig.~\ref{fig:fig1}).
    
        Having an expression for $\vec{N}$ allows the desired gate operations to be designed. The $x,y$ components of $\vec{N}$ can be chosen through the relative phase of the two pulses, which are encoded in the complex vectors $\vec{e}_{1,2}$ [\eqref{eq:e-field-time-dep}]. 
        The $z$-component $-\Delta_{\rm eff}/2$ is determined by $\mathcal{E}_{1,2}$, $\Delta_3$ and $\Delta_4$. 
        Note that terms involving the highest excited state (including $\Delta_4$) are kept, as these can have an effect on both the axis and the total angle of rotation. It is also worth noting that, as it is typical of this type of off-resonant Raman scheme, the pulse shape does not explicitly enter into the gate $R_\theta$, but rather only the area under $\xi(t)^2$. 
        
        A necessary condition for a rotation around an axis in the equator of the Bloch sphere is $\Delta_{\rm eff} = 0$, which determines $\mathcal{E}_2$ as a function of $\mathcal{E}_1$  
        \begin{equation}
            \mathcal{E}_2 = \mathcal{E}_1\sqrt{\frac{S_1}{S_2}}~,
            \label{eq:amplitude2}
        \end{equation}
        where
        \begin{equation}
            S_k = \sum_{i=5}^8 \frac{|[\bv{\mu}\cdot\vec{e}_k]_{ki}|^2}{\Delta_i}~.
        \end{equation}
        A $\theta$ rotation is then determined by (solve for $\mathcal{E}_1$ in Eq.\eqref{eq:rot-theta1}):
        \begin{equation}
            \mathcal{E}_1 = \sqrt{\frac{\theta \hbar}{2|D|\mathcal{A}} \sqrt{\frac{S_2}{S_1}} }~,
            \label{eq:amplitude1}
        \end{equation}
        in which $\mathcal{A} = \int_{-\infty}^\infty \d t\, \xi(t)^2$.
        The relative phase fixing the angle in the equator of the Bloch sphere depends on the details of the magnetic field and the polarization of the electric fields. The desired rotation around the $y$-axis can be determined numerically for each field configuration. 
        
        The terms that were neglected in the above approximations are the off-resonant cross coupling terms, which can have a significant effect on the desired gate \cite{takou_optical_2021}.
        \begin{figure}[ht!]
            \centering
            \includegraphics[width = .9\columnwidth]{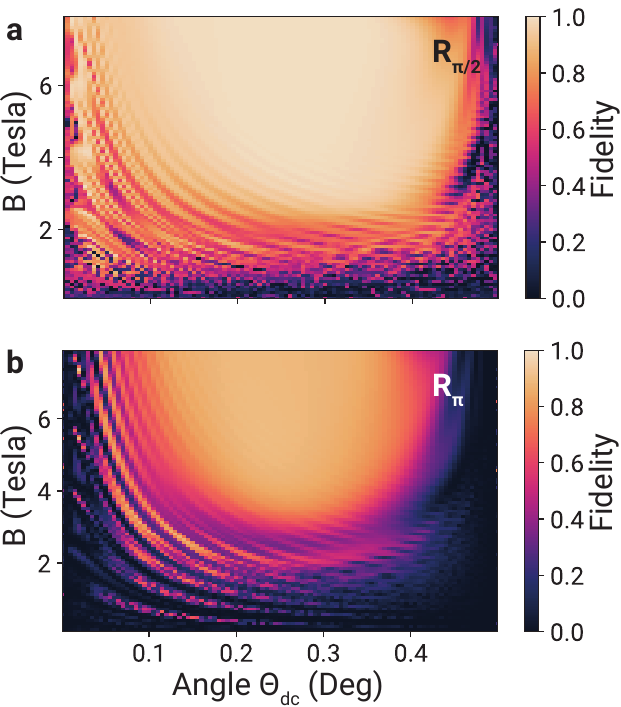}
            \caption{a) The fidelity of a $\pi/2$-rotation (x-axis, for simplicity) for a $10$ps long pulse at $\Delta_{5} = 0.1$THz as a function of the field strength and the magnetic field angle $\theta_{\rm dc}$. We chose polarization minimizing cross-coupling terms (see Appendix~\ref{appsec:rwa}). The time evolution was performed with the Floquet approximation, which shows excellent agreement with the exact time evolution. The fidelity strongly depends on the magnetic field's angle with respect to the color center's symmetry axis. A maximum is reached for $F = 0.99$ for $B = 7.36T$.
            b) Here the fidelities for a $\pi$-rotation using the same pulse parameters as in a) are shown. The fidelities reach a maximum of $F = 0.93$. The fidelities using approximate pulse parameters fall short of the highest values possible (see text).}
            \label{fig:fig2}
        \end{figure}

        We now briefly analyze the fidelity of a desired gate rotation $R_\theta$ using the pulse parameters derived in Eqs.~\eqref{eq:amplitude1} and \eqref{eq:amplitude2}. The fidelity is defined by
        \begin{align}
            F(\theta) &= |\bra{\Psi(t_i)}U^\dagger R_\theta \ket{\Psi(t_i)}|^2
            \label{eq:fidelity}
        \end{align}
        where the time evolution operator $U$ is calculated by numerically integrating  $\i \partial_t \ket{\Psi} = H(t)\ket{\Psi}$ using the control parameters derived in section \ref{section:control}, $H(t)$ given in Eq.\eqref{eq:spin_h} and $R_\theta$ in Eq.\eqref{eq:bloch_rot}. The infidelity is defined as 
        \begin{equation}
            I(\theta) = 1 - F(\theta) .
        \end{equation}
        Because the rotation gates are analyzed in the context of spin-photon entanglement, the initial state $\ket{\Psi(t_i)}$ is
        \begin{equation}
            \ket{\Psi(t_i)} = \frac{1}{\sqrt{2}}(\ket{1}\ket{e} + \ket{2}\ket{l})~.
            \label{eq:fid_in_state}
        \end{equation}
        The operators $U$ and $R_\theta$ only act on the spin qubit. 
        
        The time-dependent envelope function $\xi(t)$ [Eq.~\eqref{eq:envelope}] is assumed to be of Gaussian shape $\xi(t) = {\rm exp}(-(t-t_0)^2/4 \sigma_\tau^2) $, where we define $\tau = 2 \sigma_\tau  \sqrt{2 \log (2) }$  as the pulse duration at the full-width half-maximum with respect to the intensity of the pulses $I_k(t) \propto |\mathcal{E}_k\xi(t)|^2$. 
        The pulse shape fixes $\mathcal{A} = \sigma_\tau \sqrt{2 \pi}$ in Eqs.\eqref{eq:amplitude2}. 
    
        Fig.~\ref{fig:fig2}a, b show the $\pi/2$ and $\pi$ rotation (around the $x$ -axis) gate fidelities for $10$ps long pulses operating on the SnV ground state spin qubit using the Floquet approximation, which is in excellent agreement with the exact time evolution. Optical control of the spin state is only possible for a magnetic field that is not aligned with the symmetry axis of the color center (no strain). The polarizations were chosen so that some cross-coupling terms could be eliminated (see Appendix~\ref{appsec:rwa}). The fidelities shown for the $\pi$ rotations in Fig. \ref{fig:fig2}b require very strong magnetic fields to reach even moderate values of $F = 0.93$.
    
        It is important to note that the fidelities in Fig. \ref{fig:fig2} do not reflect an upper boundary of the attainable fidelities at a fixed detuning. Rather, they are more a reflection of the limits of the approximations that led to the calculated pulse amplitudes in Eqs.\eqref{eq:amplitude1} and \eqref{eq:amplitude2}. 
        
        We demonstrate that simplicial homology global optimization (SHGO) \cite{endres_simplicial_2018} can produce much larger gate fidelities for the coherent evolution. We optimize the fidelity of a $\pi$-rotation around the $y$-axis for a range of pulse lengths between $\tau = 10$ps and $200$ps and detunings in the range of $-100$GHz to $100$GHz at a magnetic field strength of $0.3$T, $1$T and $8$T. This range of pulse lengths is covered with spectral pulse slicing techniques \cite{boivin_spectrum_2001}, gain-switched lasers \cite{yang_development_2013} or pulse carving \cite{bekele_pulse_2018}.     
        The optimization procedure looks for the optimal magnetic field orientation, polarization, laser phase difference in case it is optimized for a $\pi/2$-rotation and pulse amplitudes -- the relative amplitudes remain fixed by \eqref{eq:amplitude2}.  We find infidelities reaching $I<10^{-7}$ for all magnetic field strengths (see Appendix~\ref{appsec:res}).
    
\begin{figure}[t!]
                \centering
                \includegraphics[width = \columnwidth]{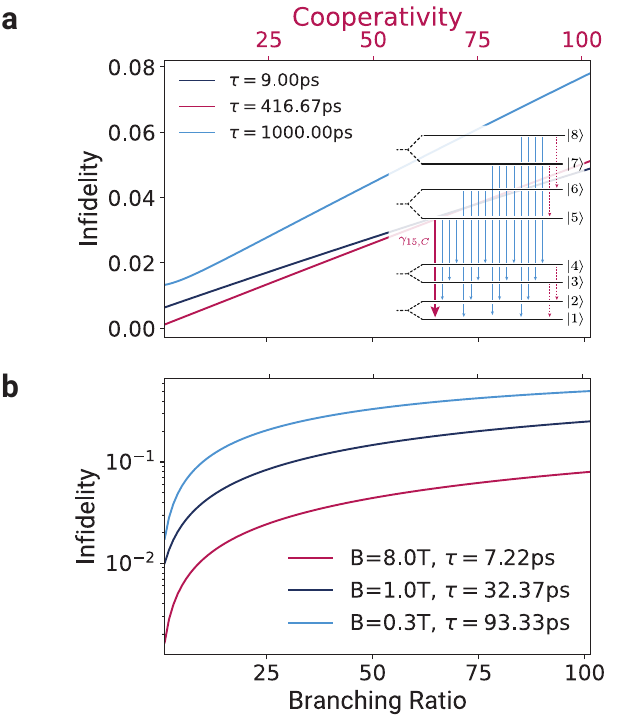}
                \caption{a) The infidelity $I(\pi)$ of a $\pi$ rotation for a range of splitting ratios/cooperativities and pulse lengths at $B = 8$ T. The pulse parameters were optimal for ($\tau = 9$ ps, $\Delta_5 = 33.33$ GHz), ($\tau = 416.67$ ps, $\Delta_5 = -41.11$ GHz) and ($\tau = 1$ ns, $\Delta_5 = -100$ GHz). The fidelity and its dependence on the branching ratio depend on the details of the optimization. For all the selected pulse lengths $I(\pi) < 8\%$ over the entire range of branching ratios. Inset: The various optical decay paths (solid, blue) and fast phononic (dashed, red) that are included in the evolution of the density matrix $\rho$ in Eq.\eqref{eq:lindblad}. The enhanced transition rate $\gamma_{15}$ \eqref{eq:gamma15purcell} is indicated by the thick (red) line. 
                b)
                Visualization of the resonant excitation infidelity increase as a function of the cooperativity for three different magnetic field strengths and at the optimized pulse lengths.
                }
                \label{fig:fig3}
        \end{figure}

        Furthermore, the rotation gate fidelities are optimized in the presence of spontaneous emission and fast non-radiative relaxation shown in Fig.~\ref{fig:fig3}a. The optical selection rules of the G4Vs depend on the magnetic field: in principle, all decay channels from the excited states to the ground states are allowed \cite{hepp_electronic_2014}. Finite branching [Eq.~\eqref{eq:branching}], in turn, creates branching errors \cite{tiurev_fidelity_2021}, which will exponentially reduce the state generation rate and quality as a function of the number of photonic qubits. The finite orbital branching ratio, which describes the ratio of rates for transitions between, for example, $\ket{5}\rightarrow\ket{1}$ and $\ket{5}\rightarrow\ket{3}$ of $\approx$ 40\% \cite{hepp_electronic_2014}, cannot be significantly influenced without a resonant structure, such as the nano-photonic crystal cavity sketched in Fig. \ref{fig:fig0}a, which Purcell-enhances transitions between $\ket{1}$ and $\ket{5}$. 
        
        The Raman gate fidelities are indirectly affected by the Purcell enhancement, due to the transient population of the excited states and subsequent spontaneous radiative emission during the interaction with the Raman laser pulses. 
        The transient population in the levels $|7\rangle$ and $|8\rangle$, which is present during the interaction of the pulses with the SnV will be subject to fast phononic decay, leading to reduced fidelities.
       
        The magnitude of the respective processes depends on the details of the magnetic field orientation and strength. Their estimates can be found in Appendix~\ref{appsec:res}. We do not include spin dephasing processes, which are assumed to be much longer for the SnV ($T_2^* > 60$ ns \cite{trusheim_transform-limited_2020}) than the pulse lengths considered here. We use the master equation in Lindblad form to incorporate spontaneous emission and fast non-radiative decay into the evolution of the system. The master equation is
        \begin{equation}
            \dot{\rho} = -\i [H, \rho] + \sum_{ij}  L_{ij} \rho L_{ij}^\dagger  - \frac{1}{2}\{ L_{ij}^\dagger L_{ij}, \rho\}    
            \label{eq:lindblad}
        \end{equation}
        where $H = H_{\rm RWA} + H_{\rm F}$ and $L_{ij} = \sqrt{\gamma_{ij}} \ketbra{i}{j}$ and $\{\cdot, \cdot\}$ denotes the anti-commutator. The various decay channels that are included in eq.\eqref{eq:lindblad} model are shown in the inset of Fig. \ref{fig:fig0}b  (see Appendix \ref{appsec:decoherence}). Fig. \ref{fig:fig3}a shows the gate fidelities for a few selected pulse lengths as a function of the branching ratio, which is defined as
        \begin{equation}
            S = \frac{\gamma_{15, C}}{\gamma_{25} + \gamma_{35} + \gamma_{45}+\gamma_{\text{ph}}}
            \label{eq:branching}
        \end{equation}
        It is assumed that the decay rate from state $\ket{5}$ to $\ket{1}$ is
        \begin{equation}
            \gamma_{15, C} = (1 + C) \gamma_{15}~,
            \label{eq:gamma15purcell}
        \end{equation}
        in the presence of a cavity with a cooperativity $C$ \cite{kimble_strong_1998}.
        As it can be seen in Fig. \ref{fig:fig3}a, the presence of spontaneous emission and fast phononic non-radiative decay has a significant impact on the optimized gate fidelities, even though the fidelities decrease by a few orders of magnitude for most pulse lengths $I < 8\%$ for $C<100$. The phononic decay has the most significant impact on the fidelities (see Fig. \ref{fig:phononvsphoton} in Appendix~\ref{appsec:res}).
    
        \begin{table}[tb]
        \centering        
        \caption{Best achieved fidelities in for three magnetic field strengths. The fidelity $F(\pi/2\circ\pi/2)$ is a $\pi$ rotation created from a composition of two $\pi/2$ rotations.}
        \begin{tabular}{cccccc}
            $B$ (T) & $\theta_{\rm dc}$ (deg) & \vspace{.1cm} $F(\pi/2)$ & $F(\pi)$ & $F(\pi/2\circ\pi/2)$ & $\tau$ (ps)\\
           \hline
           \hline
           $0.3$ & $1.0$ & $0.9953$  & $0.9831$  & $0.9906$  & $1000.00$\\
           $1.0$ & $4.5$ & $0.9916$ & $0.9830$ & $0.9833$ & $10.00$\\
           $8.0$ & $22.5$ & $0.9994$ & $0.9979$  & $0.9988$  & $416.67$\\
           \hline
           \hline
        \end{tabular}
        \label{tab:table1}
        \end{table}
        
        We summarize the highest fidelities in the presence of photonic and phononic relaxation in table \ref{tab:table1}. The table shows gate fidelities $F(\theta)$ for $\pi$ and $\pi/2$ rotations for the three magnetic field strengths $B = 0.3,\,1,\,8$ T and pulse lengths $\tau$. We also compare gate fidelities for $F(\pi)$ and the composition of two $\pi/2$ pulses $F(\pi/2\circ\pi/2)$. This composition was used in \cite{debroux_quantum_2021}, where they did not implement a single shot $\pi$ rotation. This composition creates less transient population in the higher levels, because of lower peak pulse amplitudes compared to a $\pi$ pulse. Less transient population reduces the incoherent contributions, which leads to higher fidelities in table~\ref{tab:table1}.

        In conclusion, for Raman gates, phononic spontaneous decay \cite{jahnke_electronphonon_2015} due to transient population in the far-off excited state causes the biggest impact on the fidelity. 
    
    \subsection{Excitation \label{section:excitation}}
    
        The creation of a time-bin photonic qubit, as outlined in section~\ref{section:cs_gen}, requires a high fidelity excitation pulse. Here we analyze an excitation pulse by resonantly driving the transition from state $\ket{1}$ to $\ket{5}$ with a single linearly polarized laser field
        \begin{align}
        \vec{E}(t)=  E(t) \vec{e}_{\rm L} \cos(\omega t)~,
        \end{align}
        where $E(t)$ is a Gaussian envelope, $\omega = \epsilon_5 - \epsilon_1$ and $\vec{e}_{\rm L}$ indicates the laser's polarization at the position of the color center. The objective is to find a pulse amplitude and pulse length for which the initial state
        \begin{align}
        \ket{\psi(0)}=\frac{1}{\sqrt{2}}\left(\ket{1}+\ket{2}\right)
        \end{align}
        is steered towards the target state
        \begin{align}
        \ket{\psi^{\text{tgt}}}=\frac{1}{\sqrt{2}}\left(e^{i\phi}\ket{5}+\ket{2}\right),\quad\phi\in[0,2\pi).
        \label{eq:tgtstate}
        \end{align}
        As mentioned in section~\ref{section:cs_gen}, $\phi$ is arbitrary in this context due to the interlacing of excitation pulses and $\pi$ pulses, which flip the states $\ket{1}$ and $\ket{2}$. We found that excitation fidelities did not significantly depend on the choice of polarization and hence use $\vec{e}_{\rm L} = \vec e_y$. As for the Raman rotation gates, we use a two-step procedure to optimize the rotation gates. First, a pulse amplitude is calculated using the RWA and adiabatic elimination (see Appendix \ref{sec:ex}), then the estimate is refined by scanning over varying pulse lengths to determine the highest fidelity excitation pulse duration. The pulse length is crucial here: For example long pulses are more prone to cause dephasing due to photonic decay if $\tau \approx 1/\gamma_{15,C}$ \cite{fischer_pulsed_2017, hanschke_quantum_2018}. The effect is visualized Fig. \ref{fig:fig3}b three different pulse lengths.
        We perform the optimization for three magnetic field strengths $0.3$T, $1$T and $8$T. The results are listed in table \ref{tab:table2}. 
        \begin{table}[hb!]
            \centering
            \caption{Optimized excitation pulse length and the corresponding fidelity for three different magnetic field strengths. The stronger the magnetic field, the higher the fidelity and the shorter the required pulse length.}
            \begin{tabular}{cccc}
             $B$ (T) & $ \theta_{\rm dc}$ (deg) & $\tau$ (ps) & $F$  
             \\
             \hline
             \hline
              0.3  & $1.0$ & $93.33$ & $0.9826$\\
              1    & $4.5$ & $32.35$ & $0.9900$\\
              8    & $22.5$ & $7.22$ & $0.9984$\\
              \hline
              \hline
            \end{tabular}
            \label{tab:table2}
        \end{table}

    \subsection{Single photon emission}

        Here we briefly outline the spontaneous radiative decay. We only consider single photon contributions. After a gate time $T > {\rm max}\{1/\gamma_{ij}, /1\gamma_{\rm psb}\}$, where $gamma_{\rm psb}$ is the rate of decay into the phonon sided band (PSB), the target state in Eq.~\eqref{eq:tgtstate} evolves into 
        \begin{align}
        \begin{aligned}
        \ket{\psi^{\rm tgt}} \rightarrow \quad & c_{1, \omega_{15}}\left(e^{\i\varphi_{1,1}}\ket{1}\ket{1_{\omega_{15},1}} + \ket{2}\ket{0_1}\right)
        \\ 
        &+ c_{1, 0} e^{\i\varphi_{1,0}} \ket{1} \ket{0_1}
        \\
        &+ \sum_{i=2}^8 c_{i, \omega_{15}} e^{\i \varphi_{i,1}}\ket{i}  \ket{1_{\omega_{i5},1}} 
        \\
        &+ \sum_{\omega_{\rm psb}} \sum_{i=1}^4 c_{i, \omega_{\rm psb}} e^{\i \vartheta_{i,1}}\ket{i}\ket{1_{\omega_{\rm psb},1}}, 
        \end{aligned}\label{eq:ww_emission}
        \end{align}
        where $c_{i, \omega}$ are real valued probability amplitudes, $\varphi_{i,1}$ denote the phases related to decay into the zero phonon lines at the frequencies $\omega_{i5}$ and the phases $\vartheta_{i,1}$ are related to the decay into the PSB at frequencies $\omega_{\rm psb}$, $\ket{1_{\omega, 1}}$ is a single photon state, with central frequency $\omega$, produced in a time-bin with length $T_{\rm gate}$.
        For entanglement generation, we are only interested in the probability of the desired outcome $p_{b} = |c_{1, \omega_{15}}|^2$, related to the state in the first line of Eq.~\eqref{eq:ww_emission}, and assume that unwanted events  corresponding to emission at $\omega_{i5}$ for $i>1$ and $\omega_{\rm psb}$ can be perfectly filtered. A detailed model for imperfect filtering was developed in \cite{tiurev_fidelity_2021}. We estimate the probability of a successful emission $p_{\rm b}$ from the branching ratio defined in Eq.~\eqref{eq:branching}:
        \begin{equation}
             p_{b} = 1 - \frac{1}{1 + S}
        \end{equation}
        the above probability will enter the quality measure that we define in the next section.         
        Purcell enhancement can be used to significantly increase $S$ and therefore $p_b$. Purcell enhancement, however, has to be balanced with unwanted two-photon emission events, which are more likely during the excitation process if the lifetime is shortened \cite{fischer_pulsed_2017, hanschke_quantum_2018}.
        
    \section{Multipartite Entanglement Quality Measure\label{section:cluster}}

        In the previous sections we identified spontaneous relaxation processes during the gate operations as the main obstacle to the creation of flawless states. This specifically concerns spontaneous radiative and phononic decay processes occurring during gate operations. Purcell enhancement, in particular, poses a special challenge, because it is needed for the deterministic creation of the time-bin photonic qubits. Here we construct a multipartite entanglement quality measure that allows us to account for the various imperfect gates that operate on the electronic spin during the generation of a linear cluster state or GHZ state. The measure allows us to analyze the efficacy of the generation scheme.         
        \begin{figure*}[ht!]
            \centering
            \includegraphics[width = \textwidth]{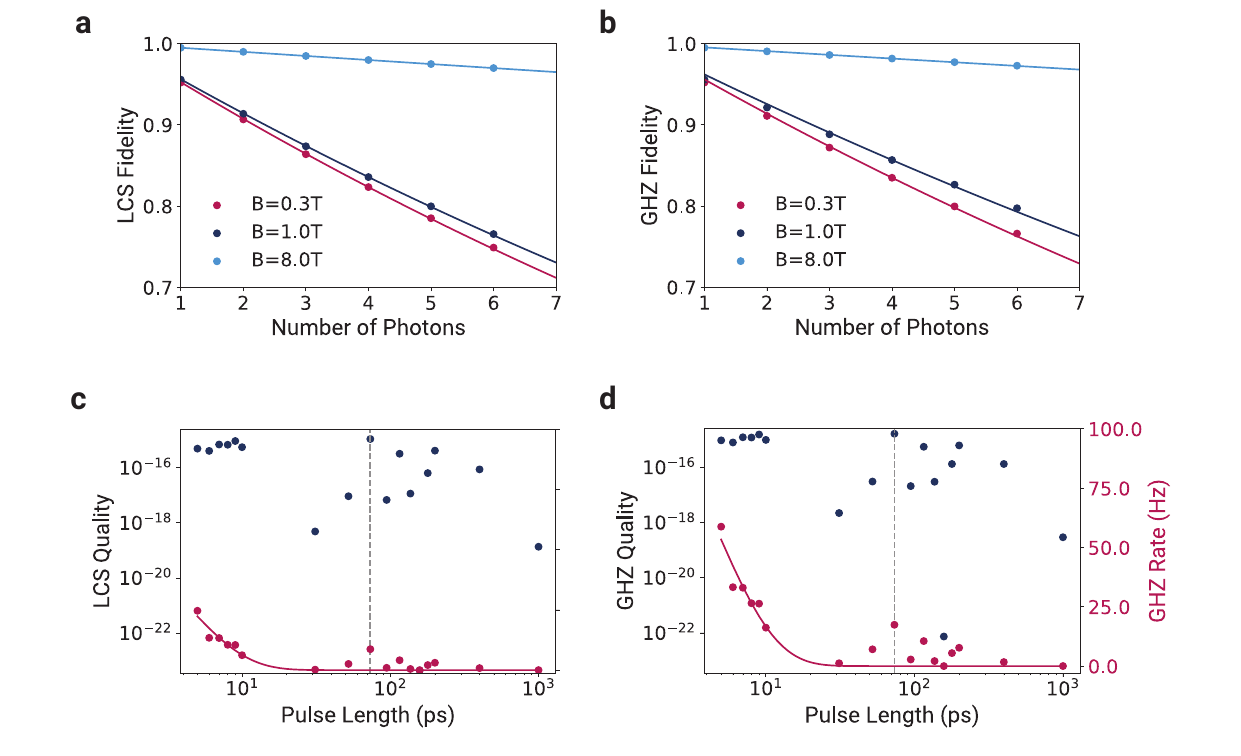}
             \caption{a) The fidelity of a LCS for an optimized branching ratio (see text) as a function of the number of photonic qubits for three field strengths. The excitation and control pulse parameters can be found in Tables~\ref{tab:table1} and \ref{tab:table2}, which informed the calculation of the depolarization errors. The fidelities are fitted with an exponential decay function for extrapolation to states with $n>6$ photons. b) The fidelities of a GHZ state for the same optimized branching ratios and pulse parameters as in a). The fidelities are again fitted with an exponential decay function. The same depolarization errors were used as in a). 
             c) LCS quality (blue dots) of a state with $100$ qubits, where we evaluated Eq.~\eqref{eq:CSqual} as a function of the Raman control pulse length $\tau$. Red dots show the state creation rate \eqref{eq:CSrate}. The quality peaks at $73.33$ps with a value of $Q_{\rm max} = 1.1\cdot 10^{-15}$ at a rate of $\Gamma({\rm LCS}, 100) = 9$Hz. A fiber attenuation length of $L_{\rm att} = 1$km was assumed. d) GHZ state quality (blue dots) of a state with $100$ qubits and rate \eqref{eq:CSrate} (red dots) as a function of the control pulse length $\tau$. The quality peaks at $73.33$ps with a value of $Q_{\rm max} = 1.7\cdot 10^{-15}$ at a rate of $\Gamma({\rm LCS},100) = 17$Hz. A fiber attenuation length of $1$km was assumed.}
            \label{fig:fig5}
        \end{figure*}
        The measure is defined by 
        \begin{equation}
            Q({\rm \mathcal{S}},n) = p_{\rm b}(n)\, p_{{\rm g}}(n)\,p_{d}(\mathcal{S}, n) \,F({\mathcal{S}},n) 
            \label{eq:CSqual}
        \end{equation}
        where $p_{\rm b}(n)$ is the probability that the desired transition from 
        $\ket{5}$ to $\ket{1}$ occurs for all spin-photon entanglement cycles (assuming perfect filtering), $p_{{\rm g}}(n)$ is the probability that no-photon is lost during the time it takes to produce the states 
        $\mathcal{S} = $ LCS, GHZ, $p_{d}(\mathcal{S}, n)$ takes into account the pure spin dephasing during the state generation, and $F({\mathcal{S}}, n)$ is the fidelity of the respective state and 
        $n$ the number of time-bin photonic qubits contained in $\mathcal{S}$. 
         
        We calculate the probability $p_{{\rm g}}(n)$ by assuming that each photonic qubit is confined to a fiber after emission (see Fig.~\ref{fig:fig0}). We find that 
        \begin{equation}
            p_{{\rm g}}(n) = \prod_{k=1}^n p_k
        \end{equation}
        where $n$ is number of time-bin qubits and
        \begin{equation}
         p_k =  \frac{\eta}{2}e^{- k  c T_{\rm tb} /L_{\rm att}}\left( 1 + e^{c T_{\rm tb} / L_{\rm att}} \right)
        \end{equation}
        where $\eta$ is the emitter to fiber coupling efficiency, $L_{\rm att} = 1$ km is the attenuation length of the fiber, $c$ is the speed of light in the fiber, $T_{\rm tb} = T_{\rm exc.} +  T_{\rm emission} + T_{\rm Raman}$, where $T_{\rm exc.}$ is the excitation duration,  $T_{\rm Raman}$ is the gate duration for the $\pi$-rotation and the $\pi/2$-rotation ($T_{\rm Raman}$ is assumed identical for both), $T_{\rm emission} = 10/\gamma_{15,C}$, which depends on the branching ratio \eqref{eq:gamma15purcell} or cooperativity respectively.
       
        The probability $p_b(n)$ is given by 
        \begin{equation}
            p_b(n) = p_{b}^{2 n} = \left(1 - \frac{1}{1 + S}\right)^{2 n}.
        \end{equation}
        We account for the spin dephasing with 
        \begin{equation}
            p_{d}(\mathcal{S}, n) = e^{- \tau(\mathcal{S},n)/\tau_{\rm c}}~,
        \end{equation}
        where $\tau({\rm LCS},n) =  2 n T_{\rm tb} + 2 T_{\rm Raman}$ and $\tau({\rm GHZ},n) =  n T_{\rm tb} + 2 T_{\rm Raman}$ is the temporal length of a state. We estimate the coherence time $\tau_{\rm c} = 1~\mu$s at $4$ K \cite{harris_coherence_2023}.
        For the calculation of $F({\mathcal{S}}, n)$, we introduce a simplified error model for estimating the depolarization of the qubit after each operation. This will help us to extrapolate $F({\mathcal{S}},n)$ for states with a large number of qubits using finite size scaling. A single depolarization error transforms the density matrix according to 
        \begin{equation}
            \mathcal{D}(\rho_R)= (1 -  \epsilon)\rho_R + \frac{\mathbb{\epsilon}}{8} {\rm tr}(\rho_R)\mathds{1}
        \end{equation}
        where $\rho_R$ is the density matrix associated to $R_\theta |\Psi(t_i)\rangle$ according to eq.\eqref{eq:fid_in_state}. After a single operation, the depolarization error and the infidelity are related by the equation $\epsilon = 8 I/7$. We use the relation between infidelity and depolarization error to map the calculated infidelities for each control operation (qubit rotation and excitation) to the depolarization error channel. A construction of the fidelity of both the LCS and GHZ states, including the branching errors, imperfect filtering, and pure phononic dephasing can be found in \cite{tiurev_fidelity_2021}.  
        
        Figs. \ref{fig:fig5}a,b show the fidelities of the LCS and GHZ state as a function of the number of time-bin photonic qubits for three magnetic field strengths. As the number of control operations increases, the fidelity drops. A magnetic field strength of $8$ T produces the best fidelities in our study, with $F({\rm LCS}, n<109) > 50\%$ and $F({\rm GHZ}, n<121) > 50\%$.  
        
        In Fig. \ref{fig:fig5}c,d we show $Q(\mathcal{S},100)$ for the LCS and GHZ states for a given control pulse length $\tau$, $B = 8$ T  and $\eta = 0.98$ \cite{bopp_sawfish_2022}. We use the optimal cooperativity for each pulse length, which is determined from the single qubit quality $Q(\mathcal{S},1)$ (see Appendix \ref{app:cs}). The quality $Q(\mathcal{S},1)$ is maximized for some finite $C$, balancing out the effects of the finite branching ratio and reduced gate fidelities, which result from the faster radiative decay. Ideally $C$ should be optimized for each photon number.     

        Fig. \ref{fig:fig5}c shows that $Q(\mathcal{S},100) > 4\cdot 10^{-16}$ for Raman control pulses with $\tau < 10$ ps with $Q_{\rm max}(\mathcal{S},100) = 1.1\cdot 10^{-15}$ for $\tau = 73.33$ ps. The probability $p_{{\rm g}}(n)$ favors shorter pulses: temporally shorter states have less time to be affected by photon loss in the fiber. The overall best quality $Q(\mathcal{S}, n)$ was achieved for $8$ T, due to the comparatively large gate fidelities (see Tables ~\ref{tab:table1},\ref{tab:table2}). We found the optimization procedure of the Raman pulses highly dependent on the pulse lengths for pulses longer than $10$ps, which produces the noisy features in $Q(\mathcal{S},100)$. Fig. \ref{fig:fig5}d shows that $Q(\mathcal{S},100) > 8\cdot 10^{-16}$ for Raman control pulses with $\tau < 10$ ps with $Q_{\rm max}(\mathcal{S},100) = 1.7\cdot 10^{-15}$ for $\tau = 73.33$ ps. 
        
        Finally, the state creation rate is given by 
        \begin{equation}
            \Gamma({\mathcal{S},n) = \frac{p_{{\rm g}}(n) p_b(n)}{\tau({\mathcal{S}},n)}}  
            \label{eq:CSrate}
        \end{equation}
        Both quality measure and rate can be seen in Fig. \ref{fig:fig5}c,d. For most pulse lengths $\Gamma({\rm LCS}, 100) > 1$Hz, $\Gamma({\rm GHZ},100) > 5$Hz. The noisy features in the rate are inherited from $Q(\mathcal{S},1)$, which determined the optimal cooperativity for a given pulse length. The quality $Q(\mathcal{S}, n)$ for the magnetic field strengths $B=0.3$T and $B=1$T is shown in Appendix \ref{app:cs}. 

        The rate and quality measure clearly favor control pulses $\tau < 80$ ps, which would be ideal for the creation of larger time-bin entangled states. Pulses cannot be too short, however, due to the quantum speed limit \cite{aifer2022quantum}.  

    \section{Discussion and Outlook \label{section:discussion}}

        In this work we studied the generation of LCS and GHZ states with negatively charged G4V. We tailored existing time-bin entanglement schemes to the G4V using the SnV as a system to benchmark the protocol. We set out to analyze the efficacy of the protocol by studying the required control operations of $\pi/2$-rotations, $\pi$-rotations, excitation and spontaneous emission.  
        
        In order to cover a broad range of magnetic fields, and in order to not be limited by the constraints of microwave control, we choose a Raman control scheme. The excitation was also modeled in the presence of relaxation processes and based on the optimal magnetic field orientation. We found that the main limiting factor of Raman control with Gaussian pulses is the transient population in the states $|7\rangle$ and $|8\rangle$, which, due to fast phononic relaxation, drastically reduce the fidelity of the control operations. For the excitation, however, it is the photonic decay.
        
        From simulations it is apparent that gate fidelities at moderate field strengths of $0.3$ T and $1$ T deteriorate due to phononic decay (see Fig. \ref{fig:phononvsphoton} in Appendix \ref{appsec:res}). A larger Zeeman splitting is useful for reducing the transient populations. We chose a magnetic field strength of  $B=8$ T to illustrate this effect. We find a $\pi$-rotation fidelity of $F(\pi) = 0.9988$ and an excitation fidelity of $F = 0.9984$ at this field strength. We find $F(LCS, 1) = 0.995$ with a quality of $0.91$ (see Fig. \ref{fig:fig4} in Appendix \ref{app:cs}). 
        
        The phononic relaxation (see Appendix \ref{appsec:decoherence}) was not modeled with the full magnetic field dependency, which would introduce more decay paths. A more complete model, such as the one proposed in \cite{harris_coherence_2023}, would help refine this shortcoming. Despite the omission, we can safely conclude that any future proposal for an optical control scheme must minimize population in $\ket{i}$ with $i>5$. This could be accomplished using, for example, a range of pulse shape optimization methods such as GRAPE \cite{petruhanov_grape_2023}, DRAG \cite{theis_counteracting_2018}, CRAB \cite{caneva_chopped_2011}, or Krotov \cite{morzhin_krotov_2019}. 
        It would also be worthwhile to compare the quality and rates in the context of microwave control \cite{rosenthal_microwave_2023, guo_microwave-based_2023, pieplow_efficient_2023}, which is unaffected by the problem of the involvement of the excited state manifold, but is also considerably slower.     
                
        The optimization of the scheme in the presence of spontaneous relaxation processes informed the expected quality of the produced state and its creation rate. We found that pulses with $\tau < 73.33$ ps produced the best quality of states, with promising rates of creation $\Gamma(\mathcal{S}, 100) > 5$ Hz. For example, for an LCS with $100$ photons the quality is maximized with $Q({\rm LCS}, 100) = 1.1\cdot 10^{-15}$ and a rate of $9$ Hz.        
        
        Our quality measure favors shorter pulses which shortens the temporal size of the state, reducing the likelihood of photon loss in a fiber. The quality and rate uniformly drop with $\eta_{\rm total}(n) = \eta^{n}$, where $\eta$ is the emitter to fiber coupling efficiency. For example, if $\eta = 0.98$ then $\eta_{\rm total}(100) \approx 0.13 $. Even small reductions in $\eta$ will cause a drastic drop in state creation rates, highlighting the importance of a near unity $\eta$.
        
        Additional important considerations that we have so far omitted from this analysis are experimental constraints with regard to the coupling efficiency of the laser fields considered when the G4V is confined in a nano-structure. These are beyond the scope of this work and should be addressed in future research investigating coupling efficiencies, which are crucial for the design of an appropriate experimental implementation. 

        Beyond a more stringent characterization of our initial exploration of Raman control, a future study should include the robustness of the presented scheme when faced with imperfections of the control fields, such as the laser fields and the magnetic field. In a very first benchmark, we determined an $\approx 4\%$ fidelity reduction for a 1\% deviation of the pulse power. 
        
        The overall conclusion from our study is that it should be possible to generate large LCS and GHZ states with $F>0.5$ using G4V even with pulse shapes that have not yet been perfectly optimized. We also conclude that state creation rates of LCS and GHZ levels with 100 photons for control pulse lengths $\tau < 73.33$ ps at $8$ T are promising indicators for successful experimental implementation.

\section*{Acknowledgements}

    The authors would like to thank Mohamed Belhassen for his scientific input. Funding for this project was provided by the European Research Council (ERC, Starting Grant project QUREP, No. 851810), the German Federal Ministry of Education and Research (BMBF, project QPIS, No. 16KISQ032K; project QPIC-1, No. 13N15858, and the Einstein Foundation Berlin (Einstein Research Unit on Quantum Devices).

\section*{Author Contributions}

    G.P compiled the model with the help of J.H.D.M and together with Y.S. performed the comprehensive numerical analysis and optimization of the control gate fidelities. M.I.M. contributed to the conception of the Raman control scheme. G.P. derived the approximations for the efficient optimization of the rotational gates and proposed the quality measure. T.S. developed the idea and supervised the project. All authors contributed to the writing of the manuscript. 

\bibliographystyle{aipnum4-1}
\bibliography{bib.bib}

\newpage
\begin{appendix}

\cleardoublepage

\section{Hamiltonian Description}\label{appsec:hamiltonian}

    Here we briefly summarize the properties of the G4V that are relevant to this work. Unperturbed, the symmetry operations leaving a G4V invariant are in the $D_{3d}$ point group. In the work by \cite{Hepp2014} an irreducible representation of the $D_{3d}$ point group was used to analyze the level structure of the G4V. We expand the G4V Hamiltonian in the same basis used in \cite{Hepp2014}:      $|e_{bx},\downarrow\rangle,\,|e_{bx},\uparrow\rangle,\,|e_{by},\downarrow\rangle,\,|e_{by},\uparrow\rangle$, where $b = g$ denotes the ground state manifold, and $b = u$ the excited state manifold. The G4V Hamiltonian then becomes
    \begin{align}
    H_{\text{G4V}}&=\begin{pmatrix}
    H^{g} & \textbf{0}\\
    \textbf{0} & H^{u}
    \end{pmatrix}~.
    \end{align}
    The Hamiltonians for the ground- and excited state manifold are
    \begin{align}
    H^{b} &=H_{A}^b+H_{SO}^b+H_{JT}^b+H_{ST}^b+H_{Z}^b~
    \end{align}
    where
    \begin{align}
    H_{A}^b &=\delta^{b}\mathds{1},\\
    H_{SO}^b &=\begin{pmatrix}
    0 & 0 & -i\lambda^{b} & 0\\
    0 & 0 & 0 & i\lambda^{b}\\
    i\lambda^{b} & 0 & 0 & 0\\
    0 & -i\lambda^{b} & 0 & 0
    \end{pmatrix},\\
    H_{JT}^b &=\begin{pmatrix}
    \Upsilon_{x}^b & 0 & \Upsilon_{y}^b & 0\\
    0 & \Upsilon_{x}^b & 0 & \Upsilon_{y}^b\\
    \Upsilon_{y}^b & 0 & -\Upsilon_{x}^b & 0\\
    0 & \Upsilon_{y}^b & 0 & -\Upsilon_{x}^b
    \end{pmatrix},\\
    H_{Z}^b &=q^{b}\gamma_L\begin{pmatrix}
    0 & 0 & i B_z & 0\\
    0 & 0 & 0 & i B_z\\
    -i B_z & 0 & 0 & 0\\
    0 & -i B_z & 0 & 0
    \end{pmatrix}\\
    &+\gamma_S\begin{pmatrix}
    B_z & B_x-iB_y & 0 & 0\\
    B_x+iB_y & -B_z & 0 & 0\\
    0 & 0 & B_z & B_x-iB_y\\
    0 & 0 & B_x+iB_y & -B_z
    \end{pmatrix}\nonumber
    \end{align}
    where $H_A$ is the unperturbed G4V Hamiltonian, $H_{SO}^b$ the spin-orbit interaction with the coupling strengths $\lambda^b$, $H_{JT}^b$ refers to the Jahn-Teller contribution with the coupling parameters $\Upsilon_x^b,\Upsilon_y^b$, and $H_Z^b$ is the Zeeman term lifting the spin degeneracy. The magnetic field is given by $\vec{B}=(B_x,B_y,B_z)$, where $\gamma_L=e/2m_e$ and $\gamma_S=2\gamma_L$, where $e$ denotes the elementary charge and $m_e$ the mass of an electron and $q^b$ is the orbital reduction factor. The parameters for the SnV are listed in table \ref{tab:parameters}.
    
    The interaction of a G4V with a classical electromagnetic field is described by
    \begin{align}
    H_{\text{Control}}(t)= -\boldsymbol{\tilde{\mu}} \cdot \vec{E}(t)
    \end{align}
    where $\vec{E}(t)$ is the time dependent electric field and the components of the transition dipole operator $\tilde{\mu}^j = -d_j\otimes \mathbb{1}$, are  
    \begin{align}
    d_x &=ae\begin{pmatrix}
    0 & 0 & 1 & 0\\
    0 & 0 & 0 & -1\\
    1 & 0 & 0 & 0\\
    0 & -1 & 0 & 0
    \end{pmatrix}~,\\
    d_y &=ae\begin{pmatrix}
    0 & 0 & 0 & -1\\
    0 & 0 & -1 & 0\\
    0 & -1 & 0 & 0\\
    -1 & 0 & 0 & 0
    \end{pmatrix}~,\\
    d_z &=2ae\begin{pmatrix}
    0 & 0 & 1 & 0\\
    0 & 0 & 0 & 1\\
    1 & 0 & 0 & 0\\
    0 & 1 & 0 & 0
    \end{pmatrix}~,
    \end{align}
    where $\mathbb{1}$ is the 2$\times$2 identity acting on the spin degree of freedom. 
    We calculate the scaling factor $a$ from the measured excited state lifetime using Fermi's golden rule in Appendix~\ref{appsec:decoherence}. 
    
    We find the eigenstates $|i\rangle$ in the main text by diagonalizing $H_{\text{SnV}}$, so that $H_{\text{SnV}}=S H_0 S^\dagger$, where $H_0 = \sum_{i=1}^8\epsilon_i \ket{i}\bra{i}$ and $S$ a basis transformation that we calculate numerically for a given $\vec{B}$. The Schr{\"o}dinger equation associated to the system is
    \begin{align}
        \ket{\dot{\psi}(t)}=-\i H(t)\ket{\psi(t)}~,
    \end{align}
    where $H(t) = H_0 + \sum_{j\rm \in\{x,y,z\}} E_j(t)\mu^j$ and $\mu^j=S^\dagger \tilde{\mu}^j S$. 
    The pulse generating Raman spin gates is given by
    \begin{align}
    \mathbf{E} (t)= \frac{1}{2}\sum_{k=1}^2 E_k(t)\left(\vec{e}_k e^{\rm i\omega_k t}+\rm c.c\right)~,
    \end{align}
    which interacts with the color center during some time $T$. The pulse shapes are described $E_k(t)$ and polarization vectors associated to the two fields are $\vec{e}_k$. Throughout this work Gaussian pulse envelopes
    \begin{align}
        E_k(t)=\mathcal{E}_k e^{-\left(t-T/2\right)^2/4\sigma^2}
    \end{align}
    are used.
    The polarizations are assumed to be linear and are parametrized in spherical coordinates:
    \begin{align}
    \vec{e}_k=G^T\begin{pmatrix}\cos(\phi_k)\sin(\theta_k)\\ \sin(\phi_k)\sin(\theta_k) \\\cos(\theta_k)
    \end{pmatrix}e^{\rm i\varphi\delta_{k2}}~,
    \end{align}
    where $\varphi$ is a phase between the first and second pulse, $\delta_{ik}$ Kronecker's delta and $G$ transforms the polarization vectors from the symmetry frame of the color center to the diamond lattice frame
    \begin{align}
    G=\begin{pmatrix}
    \frac{1}{\sqrt{6}} & -\frac{2}{\sqrt{6}} & \frac{1}{\sqrt{6}}\\
    \frac{1}{\sqrt{2}} & 0 & -\frac{1}{\sqrt{2}}\\
    \frac{1}{\sqrt{3}} &\frac{1}{\sqrt{3}} & \frac{1}{\sqrt{3}} 
    \end{pmatrix}~.
    \end{align}
    The two lowest energy eigenstates $\ket{1}$ and $\ket{2}$ form the electronic spin qubit that the Raman gates act on. 

    \section{Spontaneous Relaxation Processes} \label{appsec:decoherence}
    
    The most relevant relaxation processes are the spontaneous photonic decay and spontaneous phononic relaxation. We incorporate the relaxation processes using a master equation in Lindblad-form.
     
    The radiative decay is associated to the Lindblad-operators
    \begin{align}
        L_{ij}=\sqrt{\gamma^r_{ij}}\ket{i}\bra{j}
    \end{align}
    where $i=1,2,3,4$ and $j=5,6,7,8$. 
    We estimate the radiative decay rates using Fermi's golden rule: 
    \begin{align}
        \gamma^r_{ij}=\frac{4\alpha\omega_{ij}^3 n\vert\bra{j}\bv{\mu}\ket{i}\vert^2}{3c^2 e^2}
    \end{align}
    where $\alpha=1/137$ is the fine structure constant, $n=2.417$ the refractive index of diamond, $\omega_{ij}=\omega_i-\omega_j$ the transition frequency of level $i$ and $j$, $c$ the speed of light, $e$ the elementary charge, and $\bv{\mu}$ the transition dipole operator. 
    The transition dipole operator contains the constant $ae$. We estimate $a$ from the excited state lifetime $T_1=4.5$~ns, the SnVs DWF of 60\% at $B = 0$:
    \begin{align}
        \frac{1}{T_1}=\gamma_{32}+\gamma_{31}+\gamma_{\rm psb}
    \end{align}
    where $\gamma_{\text{psb}}=(1-DW^0)/T_1$ with $DW^0=0.6$. 
    We find $a$ by solving the above equation. 
    
    In this work we assume that the Lindblad-operators for the phononic relaxation are given by
    \begin{align}
        L_{13}=\sqrt{\gamma^p_{13}}\ket{1}\bra{3},\\
        L_{24}=\sqrt{\gamma^p_{24}}\ket{2}\bra{4},\\
        L_{68}=\sqrt{\gamma^p_{68}}\ket{6}\bra{8},\\
        L_{57}=\sqrt{\gamma^p_{57}}\ket{5}\bra{7}.
    \end{align}
    These expressions model the phononic spontaneous relaxation processes that can in principle occur between the branches ($\ket{1},\ket{2}$), ($\ket{3},\ket{4}$) and ($\ket{5},\ket{6}$), ($\ket{7},\ket{8}$). We neglect cross-coupling between even and uneven levels (e.g $\ket{1}$ and $\ket{4}$), which can, in principle, depend on the magnetic field strength and orientation. A more detailed model can be found in \cite{harris_coherence_2023}.    
    We use
    \begin{align}
    \gamma_{ij}^p=2\pi\alpha \omega_{ij}^3(n_{ij}+1)
    \end{align}
    where $\alpha=(2\pi)^{-3}\cdot 7.51\cdot 10^{-9}\text{GHz}^{-2}$ \cite{wang_transform-limited_2023}.
    Here the phononic occupation is given by
    \begin{align}
    n_{ij}=\frac{1}{e^{\hbar\omega_{ij}/k_B T-1}}
    \end{align}
    with the Boltzmann constant $k_b$ and temperature $T$.
    In this work we assume $T=0$ for calculating Raman gate fidelities, which produces adequate estimates for the SnV at $T<2$ K.  
    
    The equation of motion of the density matrix is given by the Lindblad-Master equation:
    \begin{align}
    \dot{\rho}(t)=&-\i[H(t),\rho(t)] \\ &+\sum_{(k,l)\in \mathcal{D}} L_{kl}\rho(t) L_{kl}^\dagger-\frac{1}{2}\{L_{kl}^\dagger L_{kl},\rho(t)\}
    \end{align}
    with
    \begin{align}
    \mathcal{D}&=\{(m,n)\vert m=1,2,3,4,\, n=5,6,7,8\}\\
    &\cup \{(1,3),(2,4),(6,8),(5,7)\}\nonumber.
    \end{align}
    \begin{table}[tb]
    \begin{center}
    \caption{Parameters for the Hamiltonian $H_{\rm SnV}$ \cite{Trusheim2020}}
        \centering
        \begin{tabular}{cccccccc}
         \toprule
         $b$ & $\delta^b$/THz & $\lambda^b$/GHz & $\Upsilon_x^b$/GHz & $\Upsilon_y^b$/GHz &  $q^b$\\
         $g$ & $0$ & $407.5$ & $65$ & $0$ &  $0.15$\\
         $u$ & $484.34$ & $1177.5$ & $855$ & $0$ & $0.15$\\
        \end{tabular}
        \label{tab:parameters}
    \end{center}
    \end{table}

        \begin{figure}[tb]
            \centering
            \includegraphics[width=\columnwidth]{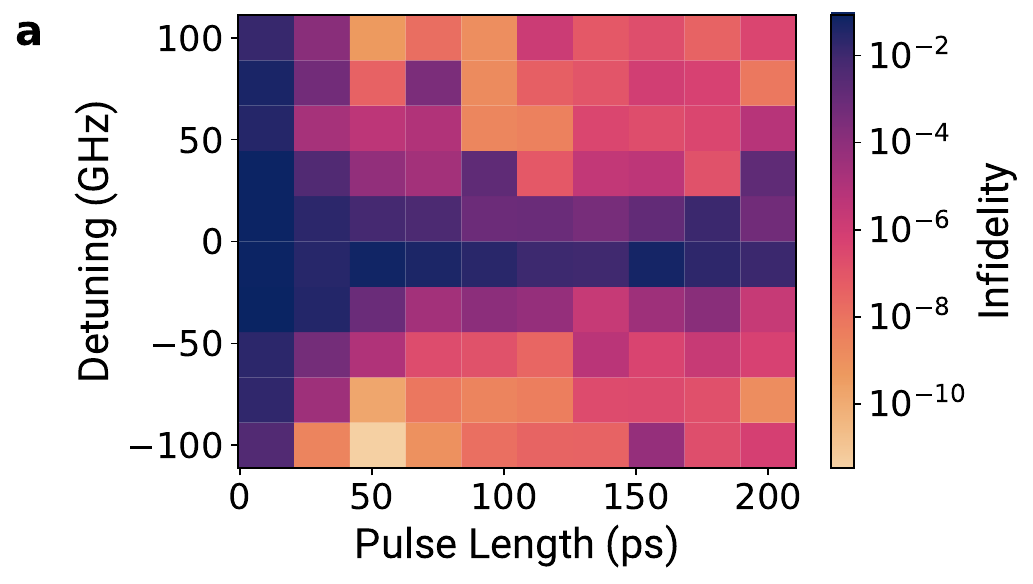}
            \includegraphics[width=\columnwidth]{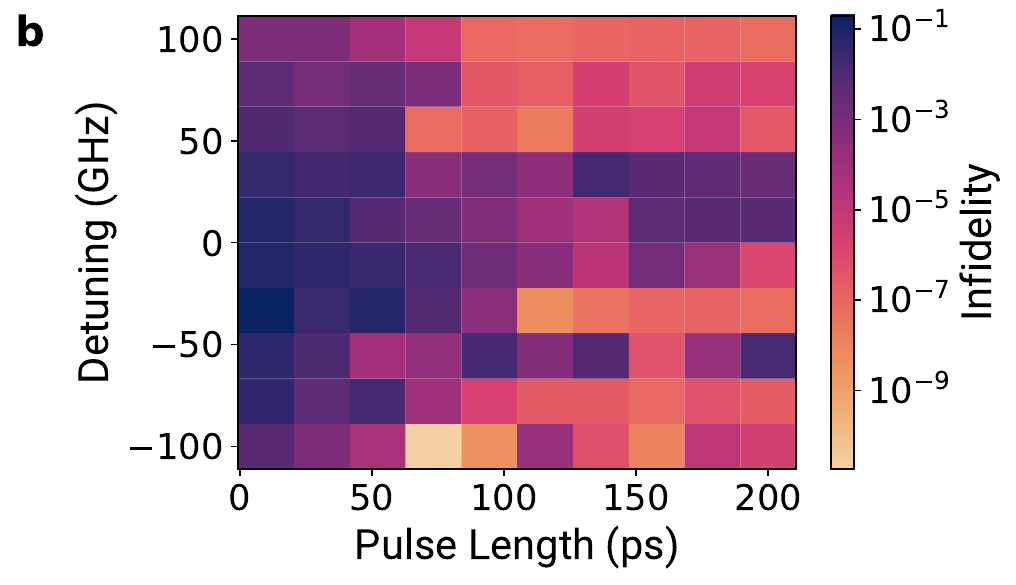}
            \includegraphics[width=\columnwidth]{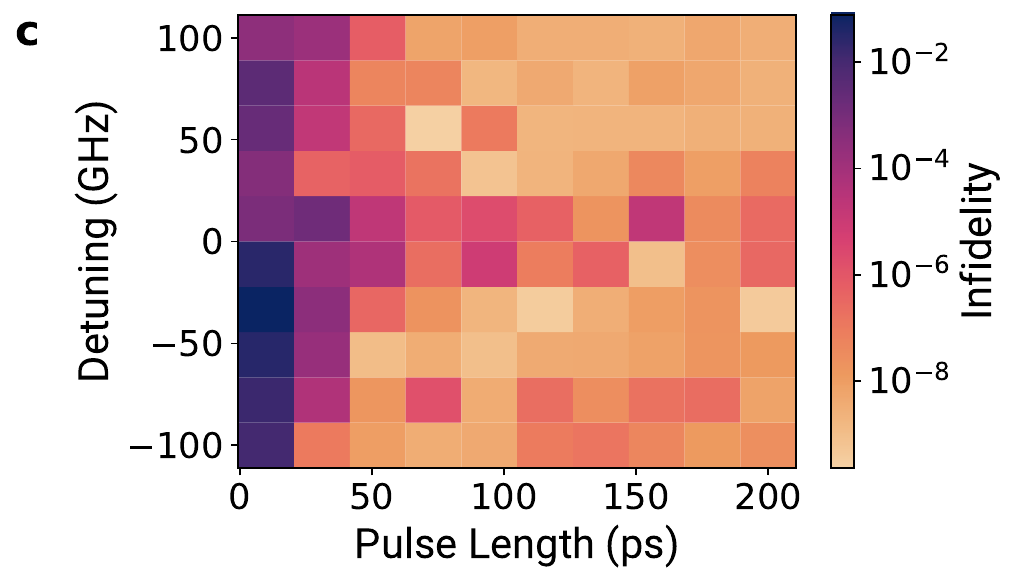}
            \caption{Evaluation of the $\pi/2$-rotation infidelity of the coherent system for optimized parameters for each detuning and pulse length for magnetic field a) $0.3$ T, b) $1$ T and c) $8$ T. Infidelities $I<10^{-7}$ are reached for some detunings and pulse lengths.}
            \label{fig:nodiss1}
        \end{figure}
        
        \begin{figure}[tb]
            \centering
            \includegraphics[width=\columnwidth]{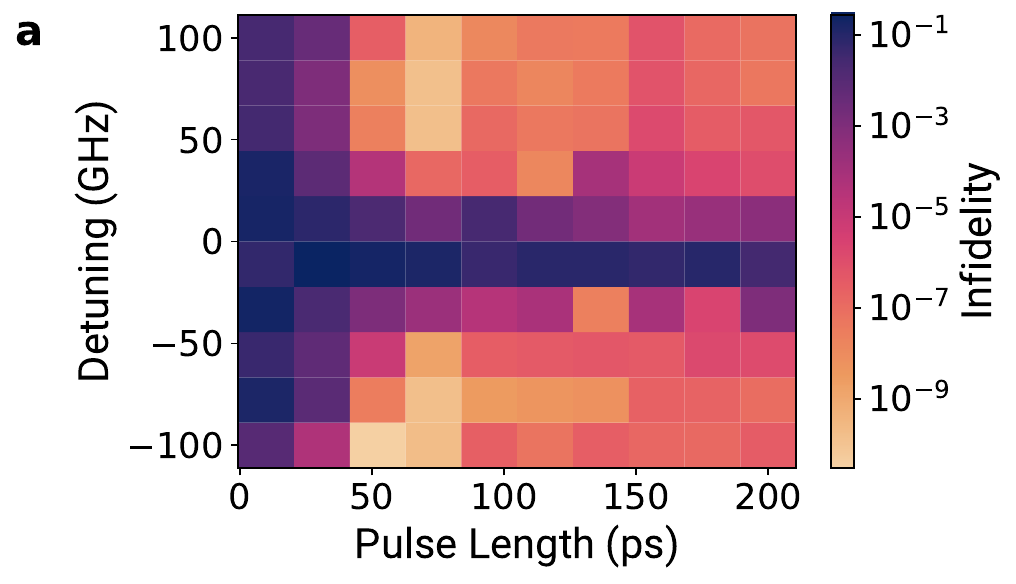}
            \includegraphics[width=\columnwidth]{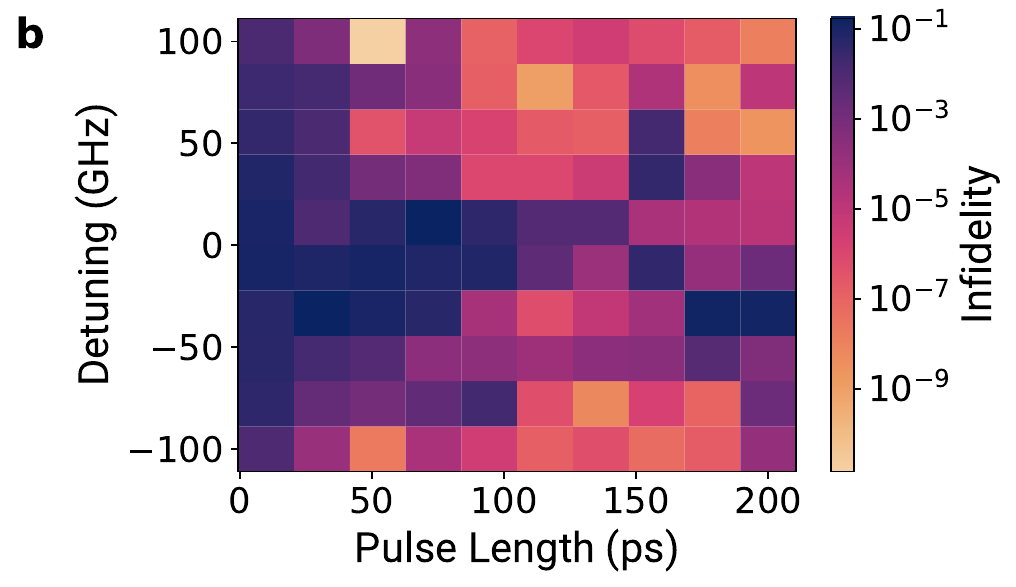}
            \includegraphics[width=\columnwidth]{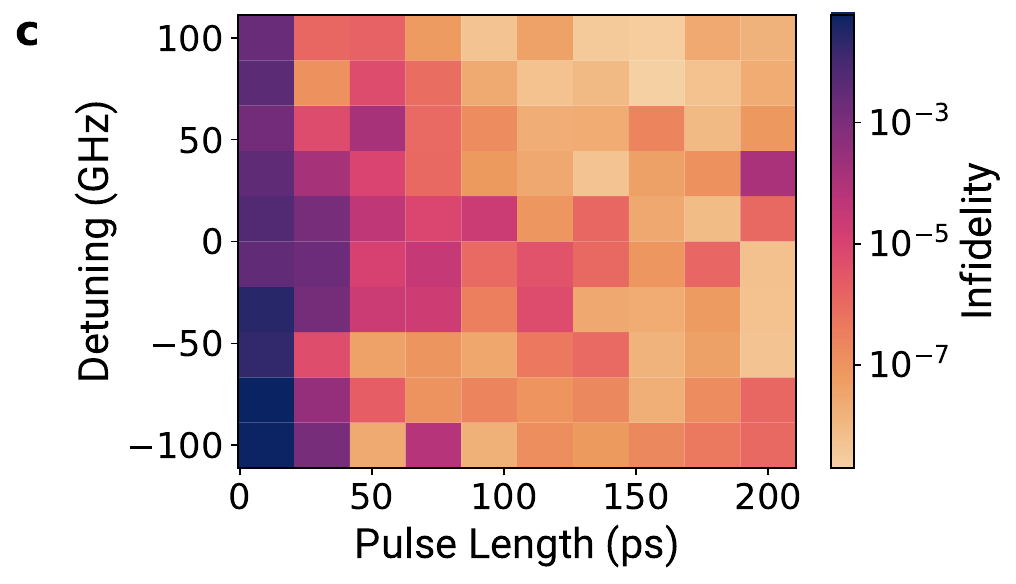}
            \caption{Evaluation of the $\pi$-rotation infidelity of the coherent system for the optimized parameters for each detuning and pulse length for magnetic field a) $0.3$ T, b) $1$ T and c) $8$ T. Infidelities $I<10^{-7}$ are reached for some detunings and pulse lengths.}
            \label{fig:nodiss2}
        \end{figure}
        
    \begin{figure}[tb]
        \begin{center}
            \includegraphics[width = \columnwidth]{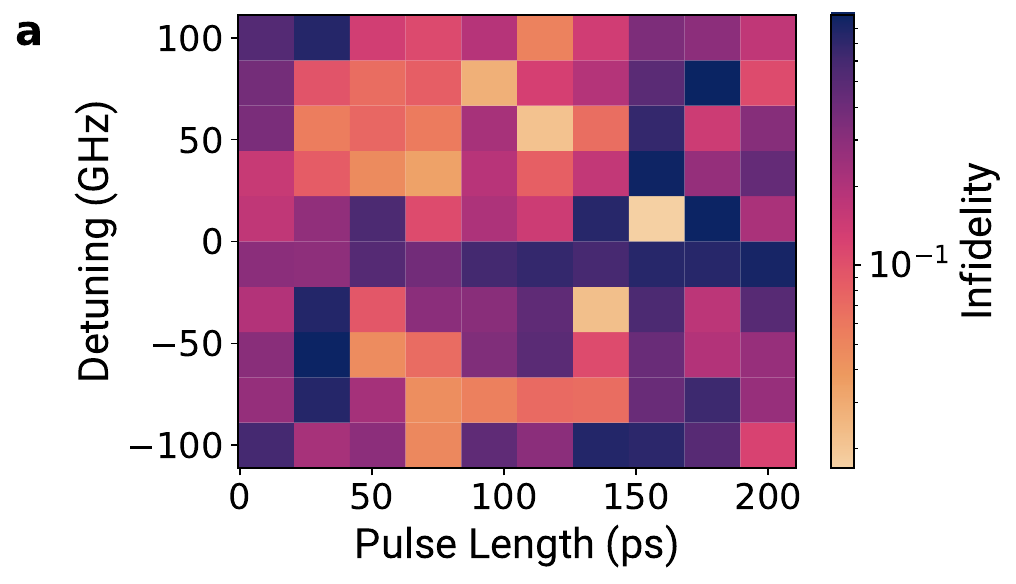}
            \includegraphics[width = \columnwidth]{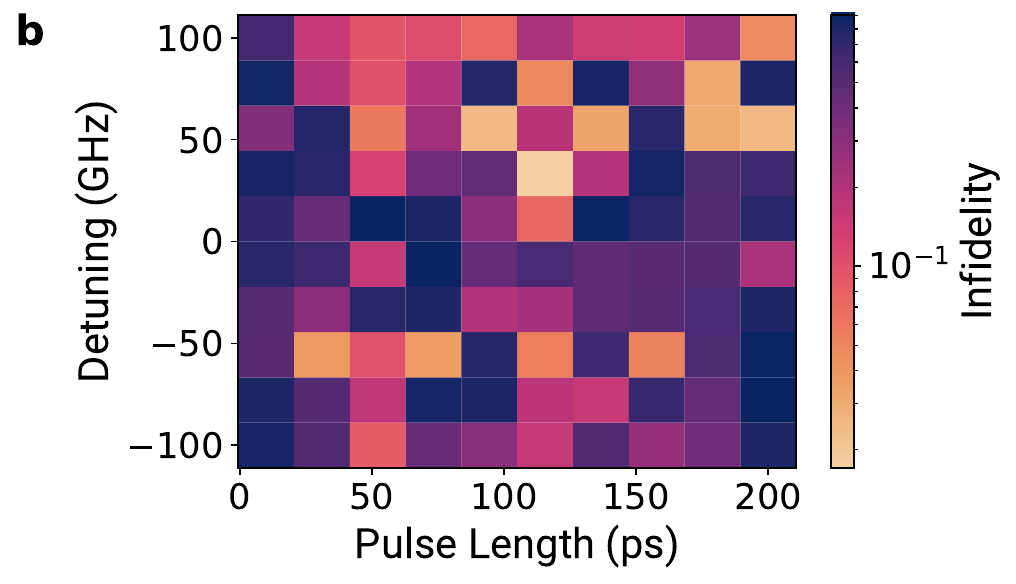}
            \includegraphics[width = \columnwidth]{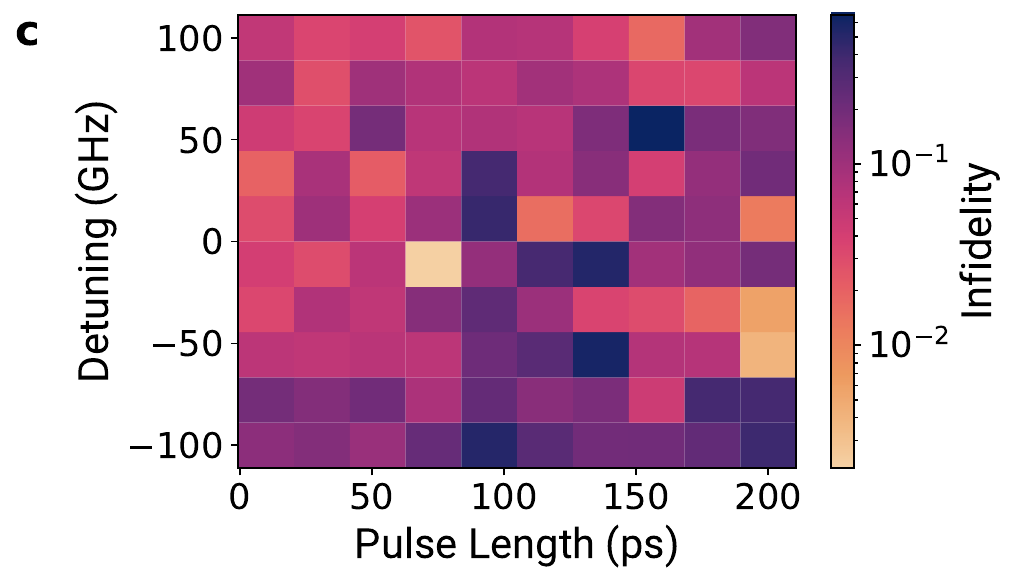}
            \caption{Evaluation of the $\pi$-rotation infidelity of the dissipative system for the optimized parameters for each detuning and pulse length $\tau$ for magnetic field a) $0.3$ T, b) $1$ T and c) $8$ T. The minimal infidelity or maximal fidelity respectively are listed in table \ref{tab:table1}.}
            \label{fig:fig11}
        \end{center}
    \end{figure}
    \begin{figure}[tb]
        \begin{center}
            \includegraphics[width = \columnwidth]{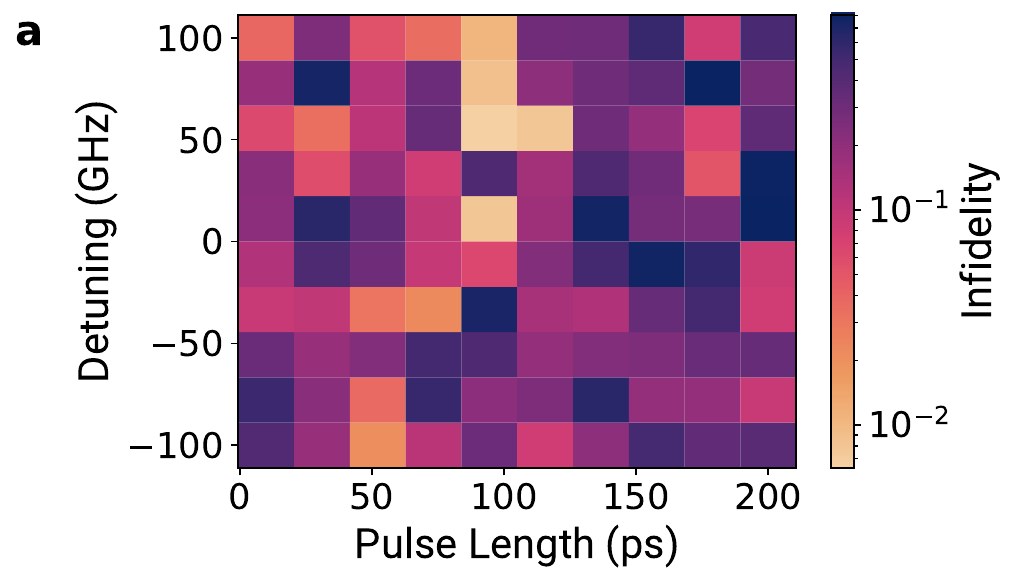}
            \includegraphics[width = \columnwidth]{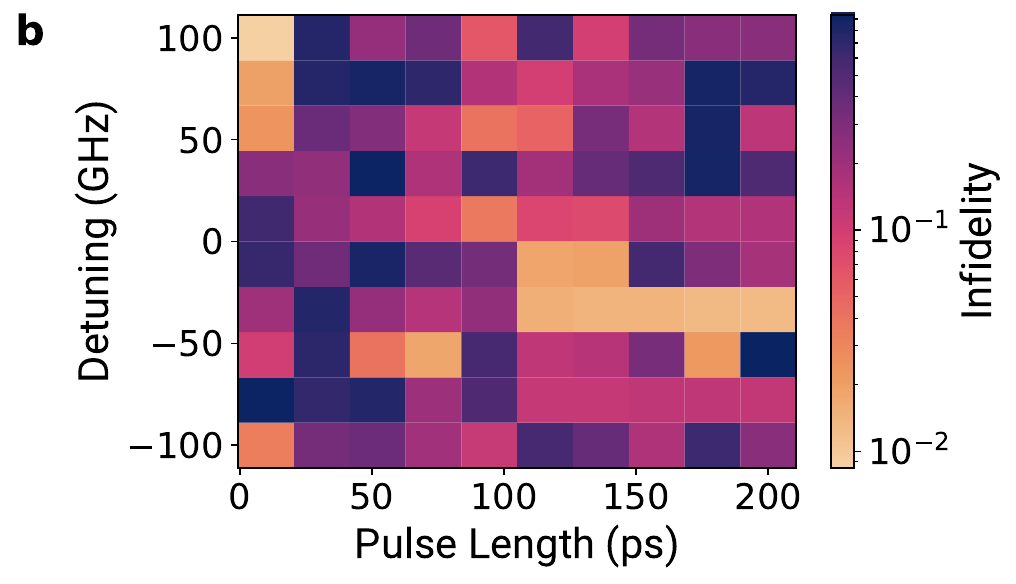}
            \includegraphics[width = \columnwidth]{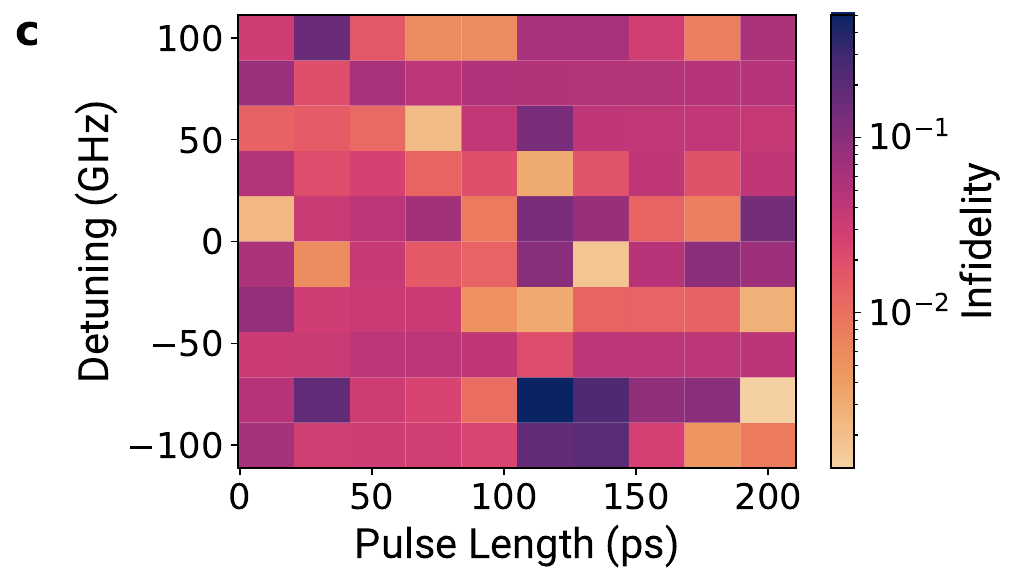}
            \caption{Evaluation of the $\pi/2$-rotation infidelity of the dissipative system for the optimized parameters for each detuning and pulse length $\tau$ for magnetic field a) $0.3$ T, b) $1$ T and c) $8$ T. The minimal infidelity or maximal fidelity respectively are listed in table \ref{tab:table1}. For $B=8$T we extended the grid beyond $200$ ps because we found the smallest value of the infidelity for $\tau = 200$ ps (extended grid not shown).}
            \label{fig:fig10}
        \end{center}
    \end{figure}
    
    \section{Floquet Approximation}\label{appsec:rwa}

    To optimize the Raman gates, the system is transformed into a rotating frame. The transformation is parameterized by the unitary operator
    \begin{align}\label{frame}
    U(t)=e^{-i\delta t}e^{-{\rm i}\sum_{j=1}^8 \xi_j t\left(2\ket{j}\bra{j}-\sum_{n=1}^8 \ket{n}\bra{n}\right)}~,
    \end{align}
    where $\delta$ and $\xi_j$ fix the frame. 
    In the rotating frame the Hamiltonian becomes 
    \begin{align}
    \tilde{H}(t)=U^\dagger (t) H(t) U(t)-{\rm i}\dot{U}(t)U^\dagger(t)
    \end{align}
    We choose $\delta$ and $\xi_j$ such that the lowest energy level is at zero energy
    \begin{align}
    \bra{1}\tilde{H}(t)\ket{1}=0\Rightarrow-\delta-\xi_1+\xi_2+...+\xi_8=0~.
    \label{eq:appph1}
    \end{align}
    We determine the other parameters from the condition 
    \begin{align}
    \text{arg}\bra{m}\tilde{H}(t)\ket{n}=0~,
    \end{align}
    where $m=1,2$ and $n=5,6,7,8$.
    This results in 
    \begin{align}
    \omega_m+2\left(\xi_m-\xi_n\right)=0~.
    \label{eq:appph2}
    \end{align}
    A solution to the simultaneous Eqs.~\eqref{eq:appph1},\eqref{eq:appph2} is given by
    \begin{align}
        &\delta =2\omega_2~,\\
        &\xi_1 =\xi_3=\frac{\omega_2-\omega_1}{2}~,\\
        &\xi_2 =\xi_4=0~,\\
        &\xi_5 =\xi_6=\xi_7=\xi_8=\frac{\omega_2}{2}~.
    \end{align}
    In the rotating frame the Hamiltonian has the form
    \begin{align}
    H'(t) =H_{\rm RWA}(t)+H_{\rm F}(t)+H_{\rm fast}(t)
    \end{align}
    where $H_{\rm RWA}$ contains the quasi time independent content of $H'(t)$, $H_{\rm F}$ is quasi time periodic oscillating with the frequency $(\omega_1-\omega_2)$ and $H_{\rm fast}(t)$ depends terms that oscillate with the frequencies $\omega_1+\omega_2$, $2\omega_1$ and $2\omega_2$. 
    For the Floquet approximation the fast oscillating terms are neglected, such that
    \begin{align}
    H_{\rm Floquet}(t) &\approx H_{\rm RWA}(t)+H_{\rm F}(t)\\
         &=H_0-\frac{E_1(t)}{2}H_1(t)-\frac{E_2(t)}{2}H_2(t)\label{eq:rwa2}
    \end{align}
    where
    \begin{align}
    H_0 &=\sum_{i=2}^8 \Delta_i\ket{i}\bra{i}
    \end{align}
    and
    \begin{align}
    &\begin{aligned}
     &H_1(t) =\sum_{\alpha\in\{x,y,z\}}e_1^\alpha\sum_{k=5}^8 \left(\vphantom{e^{-i(\omega_2-\omega_1)t}}\mu^\alpha_{1k}\ket{1}\bra{k}+\mu^\alpha_{3k}\ket{3}\bra{k} \right.
     \\
     & \left. + 
     e^{-i(\omega_2-\omega_1)t}(\mu^\alpha_{2k}\ket{2}\bra{k}+\mu^\alpha_{4k}\ket{4}\bra{k})\right)+\text{H.c.}   
    \end{aligned}
    \\
     &
     \begin{aligned}
    & H_2(t) =\sum_{\alpha\in\{x,y,z\}}e_2^\alpha\sum_{k=5}^8 
     \left(\vphantom{e^{i(\omega_2-\omega_1)t}}\mu^\alpha_{2k}\ket{2}\bra{k}+\mu^\alpha_{4k}\ket{4}\bra{k})\right.
     \\
     &\left. +e^{i(\omega_2-\omega_1)t}(\mu^\alpha_{1k}\ket{1}\bra{k}
 +\mu^\alpha_{3k}\ket{3}\bra{k})\right)+\text{H.c.}~.
    \end{aligned}
    \end{align}
    The detunings are given by
    \begin{align}
    &\Delta_2=\epsilon_2-\left(\omega_1-\omega_2\right)~,\\
    &\Delta_3=\epsilon_3~,\\
    &\Delta_4=\epsilon_4-(\omega_1-\omega_2)~,\\
    &\Delta_i=\epsilon_i-\omega_1,\quad i>5~.
    \end{align}
    The Floquet approximation shows excellent numerical agreement with the exact numerical time evolution for the studied magnetic field strengths and pulse amplitudes studied in this work.

    \section{Cross-coupling} \label{appsec:cross}

    For eliminating undesired cross-coupling terms \cite{takou_optical_2021} we made the following choice for polarization vectors:
    \begin{align}
    \begin{aligned}
        \vec{e}_1 &= (1, 0, - \mu^x_{25}/ \mu^z_{25})/\sqrt{1+|\mu^x_{25}|^2/ |\mu^z_{25}|^2}~, \\
        \vec{e}_2 &= (1, 0, - \mu^x_{16}/ \mu^z_{16})/\sqrt{1+|\mu^x_{16}|^2/ |\mu^z_{16}|^2}~.
    \end{aligned}
    \label{eq:elim-cross-talk-spin}
    \end{align}
    We used this choice for producing the results shown in Fig.~\ref{fig:fig2}.
    
    \section{Optimization}\label{appsec:res}

    The objective is to minimize the infidelity of the dissipative system. Relevant variables for the optimization are the pulse length $\tau$, the detuning $\Delta_4$, the pulse polarizations $\vec{e}_{k}$ and the pulse amplitudes $\mathcal{E}_k$ as well as the magnetic field orientation and the laser phase difference $\varphi$.
    We reduce the optimization parameter space by keeping $\mathcal{E}_1/\mathcal{E}_1 = $ constant. We choose \eqref{eq:amplitude1} and \eqref{eq:amplitude2} as start values for the pulse amplitudes for the optimization. 
    
    The minimization is numerically expensive in Liouville space. We therefore perform the optimization using the SHGO algorithm in the Hilbert space. Depending on the optimization we make a deliberate choice of the design and optimization variables. The detuning, pulse length and magnetic field strength are chosen to be design variables because they have the biggest impact of the transient population in the excited state manifold, which in turn has the biggest effect on the fidelity in the presence of dissipation. For each detuning and pulse length an optimization is performed and 
    the optimal parameters are then used to calculate the fidelity in Liouville space. We then find optimal values for the detuning and the pulse length by a simple grid optimization.
    
    The optimizations are performed for the magnetic field strengths $B=0.3,1,8\,\text{T}$ on an equidistant grid with $100$ points for $(\Delta_5,\tau)\in [-100\,\text{GHz},100\,\text{GHz}]\times [10\,\text{ps},200\,\text{ps}]$ for a $\pi$ and $\pi/2$ rotation. 
    The results without dissipation are visualized in Figs.~\ref{fig:nodiss1} and \ref{fig:nodiss2}. Under the presence of photonic and phononic relaxation processes the results visualized in Figs.~\ref{fig:fig11} and \ref{fig:fig10}. Most relevant are the highest achieved fidelities. These are listed in Table \ref{tab:table1}. It is important to mention that the fidelity primarily is deteriorated due to the phononic decay. 
    
\begin{figure}[bt]
        \centering
        \includegraphics[width=0.4\textwidth]{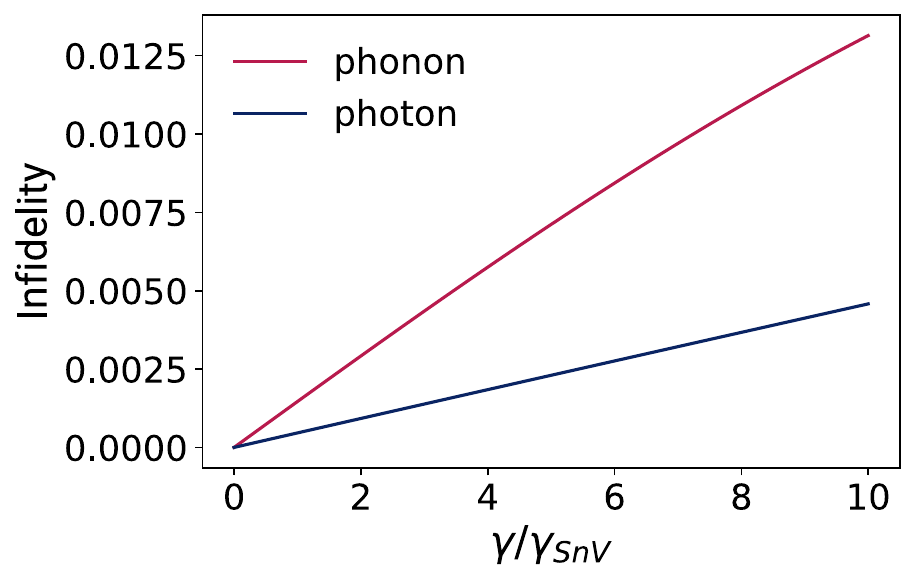}
        \caption{Comparison of the influence of the phononic and photonic decay for the opimized parameters for a Raman $\pi$ rotation and a magnetic field strength $B=8$T. The phononic decay has a more severe effect on the fidelity compared to the photonic decay. $\gamma/\gamma_{\rm SnV}$ is the ratio of the photonic (blue) and phononic (red) decay process in relation to the spontaneous emission rate of the SnV. For more details see the section~\ref{appsec:res}.}
        \label{fig:phononvsphoton}
    \end{figure}

    We compare phononic and photonic decay in \ref{fig:phononvsphoton}. We only consider spontaneous emission from level $5$ to $1$ in the range $\gamma/\gamma_{\rm SnV} \in \{0, 10\}$ ($\gamma_{\rm SnV} = 1 / T_1^{\rm SnV}$). We compare it to a phononic decay for level $7$ to $5$, without any other spontaneous emission processes. 
    We calculate the infidelity for each decay processes for the highest fidelity Raman $\pi$ pulse (see Table~\ref{tab:table1}). The result is visualized in the figure \ref{fig:phononvsphoton}.
    
    \section{Excitation}\label{sec:ex}
    
    The Floquet and RWA approximation is done in the rotating frame given by
    \begin{align}
    U(t)=\sum_{k=1}^8 e^{-i\alpha_k t}\ket{k}\bra{k}~,
    \end{align}
    where 
    \begin{align}
    \alpha_5=\omega_5,\quad \alpha_n=0\,\quad\text{for}\quad n\neq 5~.
    \end{align}
    For the Floquet approximation we eliminate terms containing $2\omega_5$. The resulting Hamiltonian is
    \begin{widetext}
    \begin{align}
    H_{\rm RWA}+H_{\rm F}(t)&=H_0+E(t)H_1(t)\\
    H_0&=\sum_{k=2}^8 \Delta_k\ket{k}\bra{k}\\
    H_1(t)&=\sum_{\alpha\in\{x,y,z\}} e_{\rm L}^{\alpha}\left(\sum_{m\neq 5} \mu_{m5}^\alpha\ket{m}\bra{5}+\sum_{m,n\neq 5} e^{i\omega_5 t} \mu_{mn}^\alpha\ket{m}\bra{n}\right)+\rm H.c.
    \end{align}
    \end{widetext}
    with $\Delta_k=\epsilon_k$ for $k\neq 5$ and $\Delta_5=0.$
    
    We estimate the excitation pulse amplitude by neglecting $H_{\rm F}$ and thereby performing the RWA:
    \begin{align}
    H_{\rm RWA}(t)=\sum_{i=1}^8 \Delta_i\ket{i}\bra{i}+E(t)\sum_{i=1}^8 c_i\ket{i}\bra{5}+\rm H.c. 
    \end{align}
    with the coefficients $c_i=\sum_{\alpha\in\{x,y,z\}} e_{\rm L}^\alpha\mu_{i5}^\alpha$.
    After adiabatic elemination we find
    \begin{align}
    \tilde{H}(t)&=\begin{pmatrix}
    0 & c_1 E(t)\\
    c_1^* E(t) & \sum_{i\neq \{5,1\}} \frac{1}{\Delta_i}\vert c_i\vert^2 E^2 (t)
    \end{pmatrix}\\
    &\approx\begin{pmatrix}
    0 & c_1 E(t)\\
    c_1^* E(t) & 0
    \end{pmatrix}
    \end{align}
    due to large detunings $\Delta_i$ for $i\neq 1,5$ the last approximation is adequate. 
    With $E(t)=\mathcal{E}e^{-\left(t-T/2\right)^2/4\sigma^2}$ the propagator at the time $T$ is
    \begin{align}
    U(T,0)=\exp\left\{i\theta\sigma\cdot n\right\}=\cos(\theta)\mathds{1}+i\sin(\theta)\sigma\cdot n
    \end{align}
    with $n=1/\vert c_1\vert (\text{Re}(c_1),\text{Im}(c_1),0)^T$, $\theta=-\vert c_1\vert\int_{0}^T E(t)\approx-\vert c_1\vert\int_{\mathbb{R} }E(t)\,\text{d}t=-2\vert c_1\vert\sigma\sqrt{\pi}\mathcal{E}$ and $\sigma=(\sigma_x,\sigma_y,\sigma_z)^T$. The pulse amplitude $\mathcal{E}$ is computed such that 
    \begin{align}
    \ket{\psi^{\text{tgt}}}=U(T,0)\ket{\psi(0)}
    \end{align}
    holds for $\ket{\psi(0)}=\ket{0}$ and $\ket{\psi^{\text{tgt}}}=e^{{\rm i}\phi}\ket{1}$ for some phase $\phi\in [0,2\pi)$. It is
    \begin{align}
    U(T,0)\ket{\psi(0)}&=\cos(\theta)\ket{0}\\
    &+\frac{\rm i}{\vert c_1\vert}\sin(\theta)\left(\text{Re}(c_1)-{\rm i}\text{Im}(c_1)\right)\ket{1}\\
    &=e^{{\rm i}\phi}\ket{1}
    \end{align}
    which implies 
    \begin{align}
    -\theta=2\vert c_1\vert\sigma\sqrt{\pi}\mathcal{E}=\frac{\pi}{2}.
    \end{align}
    The targeted pulse amplitude thus is
    \begin{align}
    \mathcal{E}=\frac{\sqrt{\pi}}{4\sigma\vert c_1\vert}.
    \end{align}
    \section{Cluster and GHZ State Quality Measure}\label{app:cs}
\begin{figure*}[tb]
        \centering
        \includegraphics[width = \columnwidth]{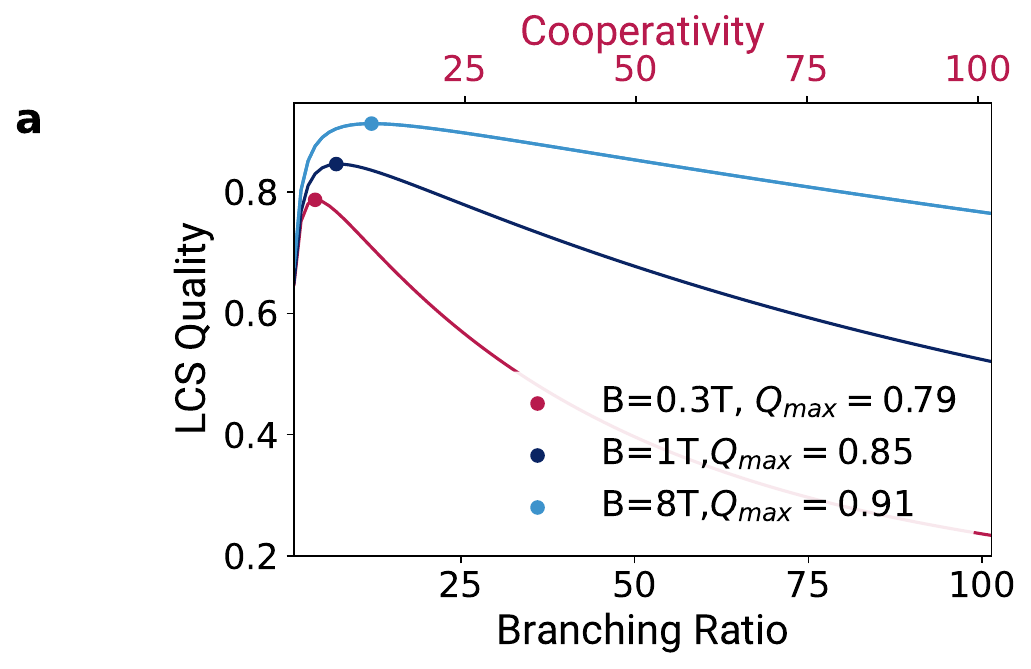}
        \includegraphics[width = \columnwidth]{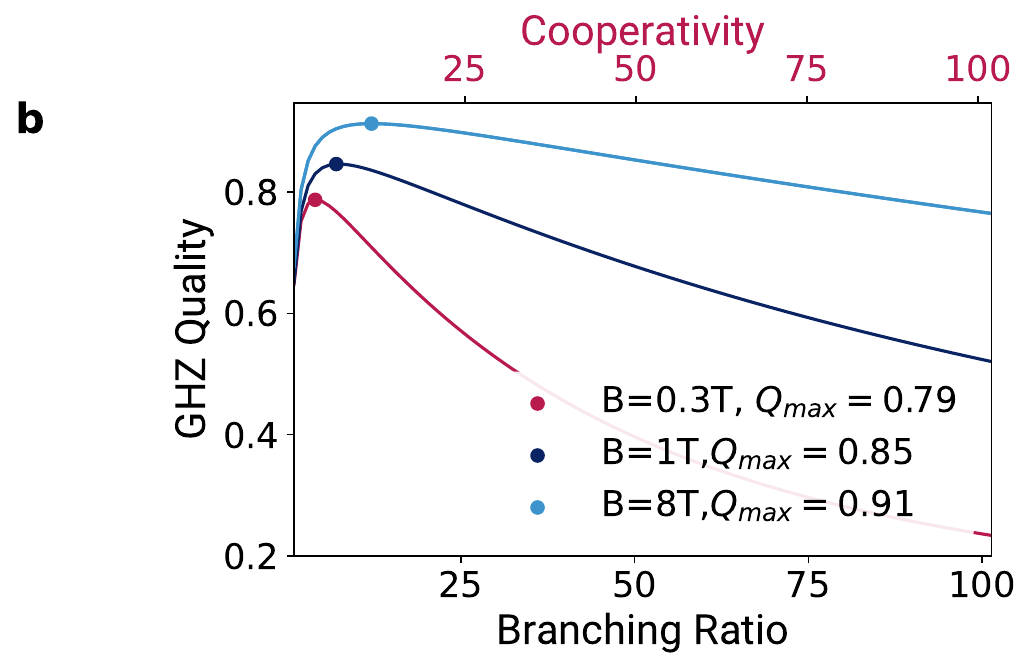}
        \caption{
         a) LCS quality measure for a single photon as a function of the branching ration and a fiber attenuation length $L_{\rm att}=1$ km. We choose the best control pulse parameters for each field strength (see Tables.~\ref{tab:table1} and \ref{tab:table2}.). 
         b) GHZ quality measure for a single photon as a function of the branching ration and a fiber attenuation length $L_{\rm att}=1$ km. We choose the best control pulse parameters for each field strength (see Tables.~\ref{tab:table1} and \ref{tab:table2}.)}
        \label{fig:fig4}
    \end{figure*}

    The quality measure dependence on the cooperativity or branching ratio respectively for a single photon LCS and GHZ state is illustrated in Fig. \ref{fig:fig4}.
    \begin{figure*}[tb]
        \centering
        \includegraphics[width=\columnwidth]{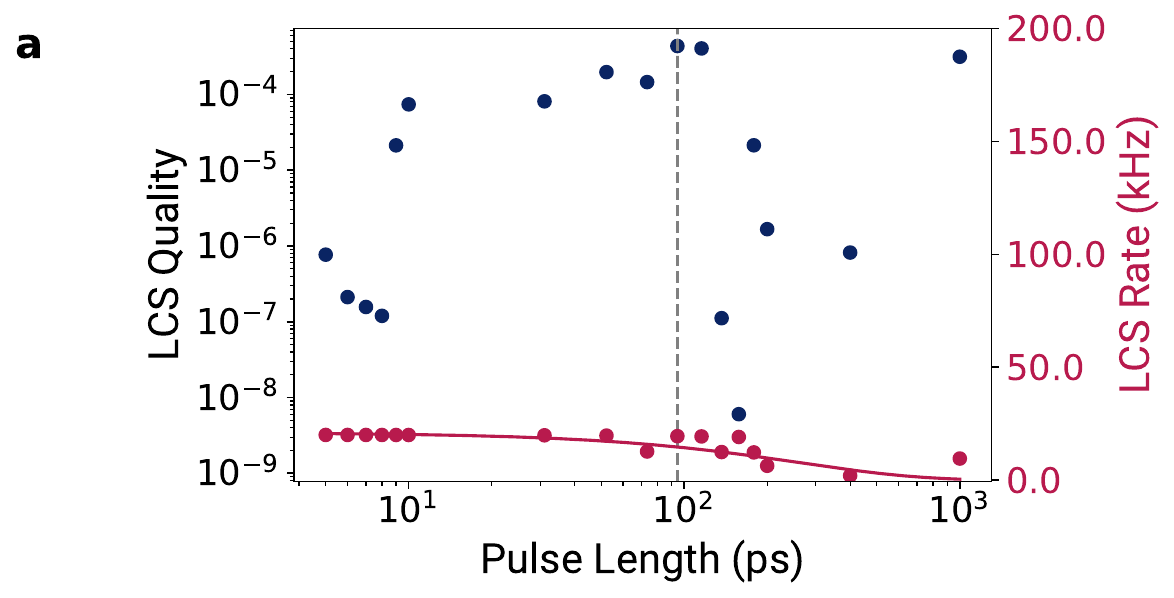}
        \includegraphics[width=\columnwidth]{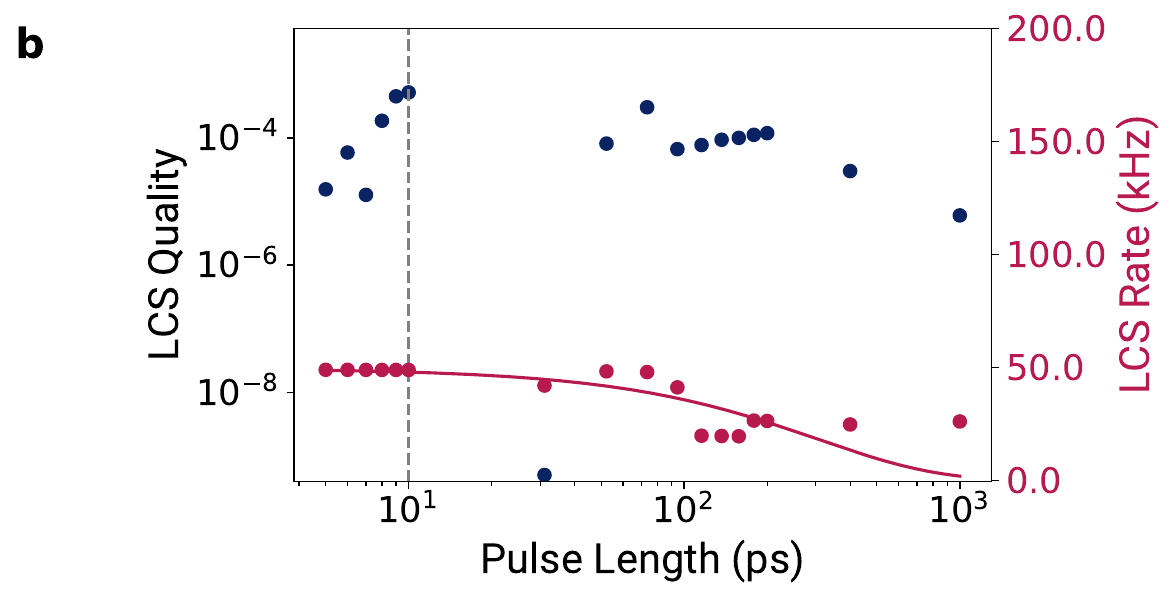}
        \caption{Cluster state quality of a cluster state with $20$ qubits \eqref{eq:CSqual} and \eqref{eq:CSrate} as a function of the pulse length. The quality peaks at a) $94.44$ps with a value of $Q_{\rm max} = 7.1\cdot 10^{-5}$ for $B=0.3$T at a rate of $\Gamma_{\rm CS} = 20$kHz. A fiber attenuation length of $1$km was assumed. b) The peak is at $10$ps for $B=1$T and $Q_{\rm max} = 5\cdot 10^{-4}$ with $\Gamma_{\rm CS} = 49$kHz.}
        \label{fig:lcs1}
    \end{figure*}
       \begin{figure*}[tb]
        \centering
        \includegraphics[width=\columnwidth]{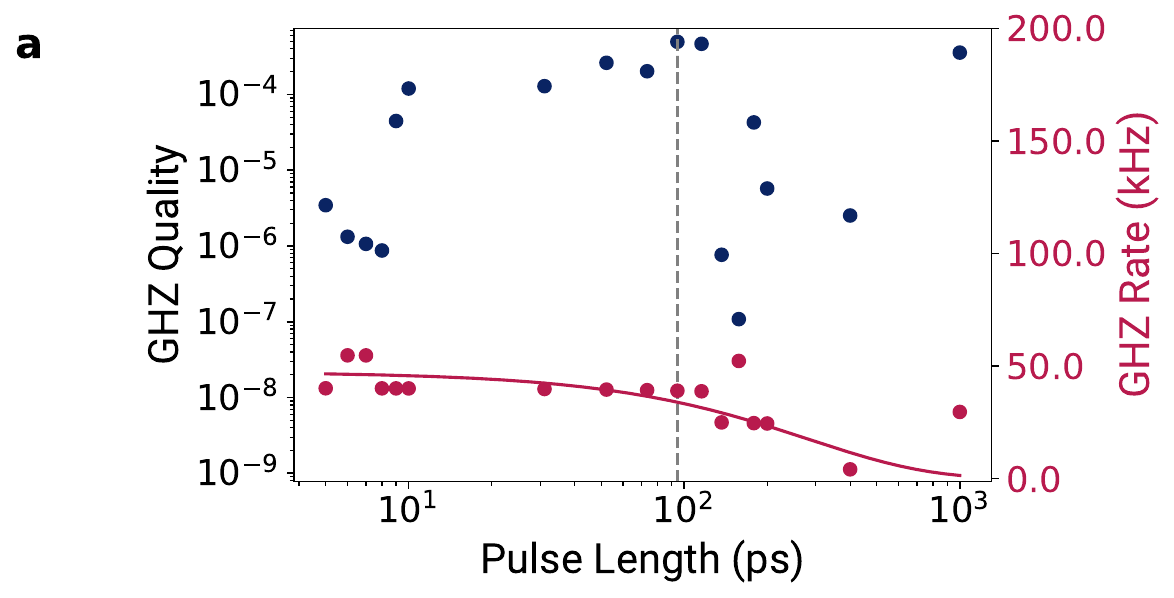}
        \includegraphics[width=\columnwidth]{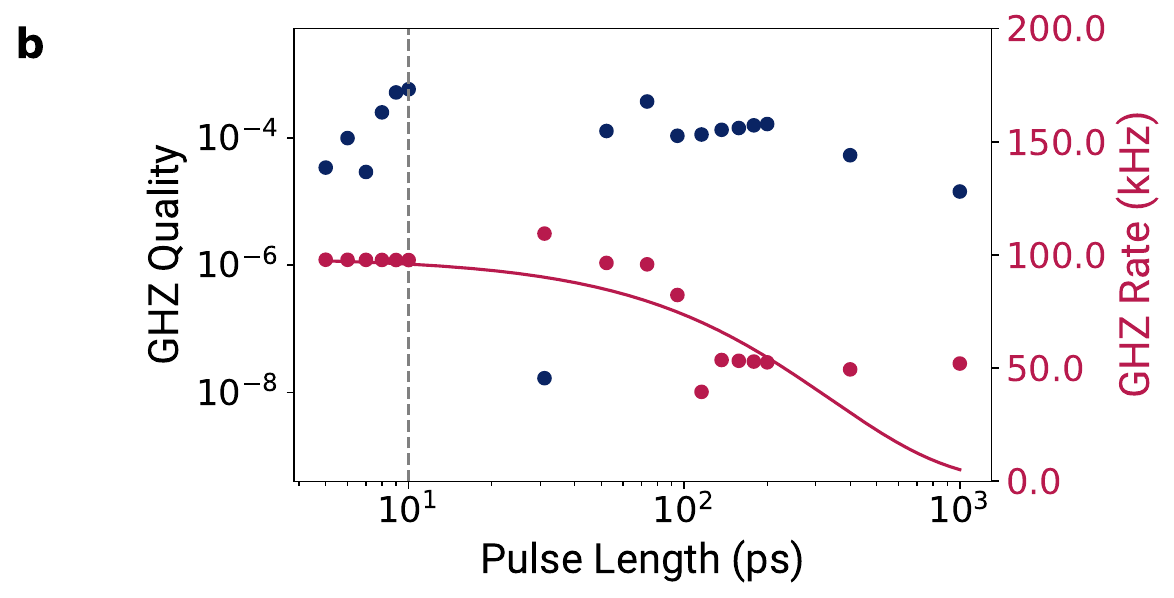}
        \caption{GHZ quality with $20$ qubits \eqref{eq:CSqual} and \eqref{eq:CSrate} as a function of the pulse length. The quality peaks at a) $94.44$ps with a value of $Q_{\rm max} = 5\cdot 10^{-4}$ at a rate of $\Gamma_{\rm CS} = 39$kHz for $B=0.3$T. A fiber attenuation length of $1$km was assumed. b) The peak is at $10$ps for $B=1$T and $Q_{\rm max} = 5\cdot 10^{-4}$ with $\Gamma_{\rm CS} = 97$kHz.}
        \label{fig:lcs2}
    \end{figure*}
    Figs.~\ref{fig:lcs1} and \ref{fig:lcs2} show the LCS and GHZ quality for the magnetic field strengths $B=0.3$T and $B=1$T. We choose $n=20$ photons because the qualities are too low for $n=100$. 
    \end{appendix}
\end{document}